\documentclass{aa}  

\usepackage{graphicx}
\usepackage{txfonts}
\usepackage{lipsum}
\usepackage{subcaption}         
\usepackage{lscape}             
\usepackage{placeins}           

\usepackage[sticky-per=false,
separate-uncertainty,
angle-symbol-over-decimal,
print-unity-mantissa=false
]{siunitx}
\usepackage{xspace}
\newcommand{\mdust}{$M_\mathrm{dust}$\xspace}
\newcommand{\mstar}{$M_\mathrm{stars}$\xspace}

\newcommand{\halpha}{H$\alpha$\xspace}
\newcommand{\hbeta}{H$\beta$\xspace}

\newcommand{\singlet}[3]{[#1\,{\sc #2}]\,$\lambda #3$}

\newcommand{\logoh}{\ensuremath{12+\log(\mathrm{O}/\mathrm{H})}}

\DeclareSIUnit\jansky{Jy}
\DeclareSIUnit\erg{erg}
\DeclareSIUnit\mag{mag}
\DeclareSIUnit\parsec{pc}
\DeclareSIUnit\arcsec{arcsec}
\DeclareSIUnit\arcsect{^{\prime\prime}}
\DeclareSIUnit\msun{M_\odot}
\DeclareSIUnit\lumsol{L_\odot}
\DeclareSIUnit\zsun{Z_\odot}
\DeclareSIUnit\year{yr}
\DeclareSIUnit\beam{beam}
\DeclareSIUnit\angstrom{\text {Å}}

\defcitealias{Popping2017DustEvolutionSam}{P17}
\defcitealias{Parente2023CosmicDustSemiAnalytic}{P23}
\defcitealias{RagoneFigueroa2024MuppiSim}{R24}

\usepackage[
colorlinks=true,
linkcolor=blue,
citecolor=blue,
filecolor=blue,
urlcolor=blue,
unicode=true
]{hyperref}

\begin{document}

\title{ALMA Chemical Evolution (ACE) survey: the dust content of subsolar metallicity galaxies at cosmic noon}


%

\author{
	M. Solimano\inst{1}\corrauth{cosmo.manus@gmail.com}
  \and I. Shivaei\inst{1}
  \and G. Popping\inst{2}
  \and L. A. Boogaard\inst{3}
  \and I. Langan\inst{1}
  \and R. Popescu\inst{4}
  \and N. N. Geesink\inst{2}
  \and A. Pope\inst{4}
  \and L. Arriscado\inst{1, 5}
  \and B. Mobasher\inst{6}
  \and D. Narayanan\inst{7, 8}
  \and M. Parente\inst{7}
  \and N. Reddy\inst{6}
  \and R. Sanders\inst{9}
}

\institute{Centro de Astrobiolog\'ia (CAB), CSIC-INTA, Ctra. de Ajalvir km 4, Torrej\'on de Ardoz, E-28850, Madrid, Spain
	\and European Southern Observatory, Karl-Schartzschild-Str. 2 D-85748, Garching, Germany
	\and Leiden Observatory, Leiden University, PO Box 9513, NL-2300 RA Leiden, The Netherlands
	\and Department of Astronomy, University of Massachusetts, Amherst, MA 01003, USA
	\and Departamento de F\'isica de la Tierra y Astrof\'isica, Facultad de Ciencias F\'isicas, Universidad Complutense de Madrid, E-28040, Madrid, Spain
	\and Department of Physics and Astronomy, University of California, Riverside, 900 University Ave, Riverside, CA 92521, USA
	\and Department of Astronomy, University of Florida, 211 Bryant Space Sciences Center, Gainesville, FL 32611 USA
	\and Cosmic Dawn Center at the Niels Bohr Institute, University of Copenhagen and DTU-Space, Technical University of Denmark
	\and Department of Physics and Astronomy, University of Kentucky, 505 Rose Street, Lexington, KY 40506, USA
 }

\abstract{
	Dust plays a key role in galaxy evolution by influencing star formation and shaping the observed spectrum of galaxies. 
However, at $z\sim2$ (cosmic noon) our knowledge of the dust mass budget is currently limited to the most massive, metal-rich systems, which are not representative of the bulk galaxy population.
Here, we probe the lower mass, subsolar metallicity regime by measuring the dust mass of 25 galaxies at $z\approx 2.3$ from the ALMA Chemical Evolution (ACE) Large Program.
The sample contains star-forming galaxies in the COSMOS field with robust strong-line metallicities down to $\sim 0.3 Z_\odot$.
Using the dust continuum emission detected at $873\,\mu\mathrm{m}$ or $1.3\,\mathrm{mm}$ we constrain the dust mass by assuming an optically thin single-temperature modified blackbody (MBB).
The resulting masses of the detected galaxies have a median of  $M_\mathrm{dust}=6.4\times10^{7}\,\mathrm{M}_\odot$, which is four  to nine times larger than those of $z=0$ galaxies at matched metallicity and stellar mass, but more than ten times lower than IR-selected galaxies at $z\approx 2$.  Under a fixed set of MBB parameters, we find positive correlations between dust mass and stellar mass, metallicity, and star formation rate (SFR).
In contrast, we find that the $M_\mathrm{dust}/M_\mathrm{stars}$ (DtS) ratio scatters around $10^{-2}$ but no clear correlation with metallicity, unless we adopt a metallicity-dependent prescription for dust temperature.  We also find a positive correlation between $M_\mathrm{dust}/\mathrm{SFR}$ and metallicity that suggests that dust buildup becomes more efficient at higher metallicities, likely due to the role of ISM grain growth or enhanced dust survival. We also recover the well known relation between  DtS ratio and specific SFR (sSFR), without evidence of a turnover at high sSFR that would signal inefficient dust buildup. Finally, we measure a large molecular gas fraction among ACE CO(3-2)-detected galaxies ($M_\mathrm{mol}/M_\mathrm{stars} \gtrsim$ 1)  that potentially explains both the sSFR and DtS excess over $z\approx0$ galaxies.
}


\maketitle
\nolinenumbers

\section{Introduction}\label{sec:intro}
A fraction of the metals in the interstellar medium (ISM) of galaxies is condensed in the form of dust grains.
Despite accounting for less than 0.1\% of the mass of a galaxy, dust plays a central role in galaxy evolution: first, it  participates in the process of star formation by enabling the creation of molecules and providing a cooling channel for the interstellar gas \citep[e. g.,][]{Draine2011IsmBook}. For this reason, dust is often used as a tracer of cold molecular gas \citep[e. g.,][]{Hildebrand1983DustCloudMassFromThermalEmission, Eales2012DustEmissionMassIsm, Carilli&Walter2013Review, Groves2015DustEmissionTracesGas, Scoville2016MainSequenceAndGasMass}. Second, dust shapes the spectrum of galaxies by attenuating the ultraviolet and optical starlight, and then reprocessing the absorbed energy into infrared (IR) emission \citep[e. g.,][]{Galliano2018IsmDustNearbyReview}.

The total amount of dust in a galaxy is set by the balance between dust creation and destruction processes.
Dust mass increases via grain production in stellar sources such as supernovae \citep[e. g.,][]{Gall2011ProductionOfDustMassiveStarsHighz, Sarangi2018DustSupernovaeFormation} or asymptotic giant branch stars \citep[e. g.,][]{HofnerAndOlofsson2018MassLossAgbStars}, and via metal accretion into existing dust grains, in a process known as interstellar dust growth \citep[e. g.,][]{Barlow1978IsmDustGrowth}.
The opposite process is the removal of metals from dust grains (sputtering).
This occurs when dust is embedded in a turbulent hot gas \citep[e. g.,][]{McKee1987ShocksAndGrainDestruction, Tielens1994GasGrainSputteringIsm}  or exposed to energetic cosmic rays \citep[e. g.,][]{Draine1979DustDestructionMechanisms}.
Dust mass also decreases when grains get locked into forming stars (astration), or when they are blown away by galaxy-scale outflows \citep[e. g.,][]{Feldmann2015OutflowsDriveLowDust, Nanni2020DustEvolLowMetallicity}.
An observational characterization of the dust content and its relation to gas-phase metals in galaxies is therefore crucial to understand the relative importance of each mechanism.

In this context, several studies have quantified the dust mass in nearby galaxies and found multiple and complex trends with other
galactic properties such as stellar mass, star formation rate (SFR), gas mass, and metallicity.
For example, \citet{deLooze2020JingleIVDustScaling} find that dust mass scales with the stellar mass, although with a sublinear slope and a \num{0.39} dex scatter.
Their results imply that at $z\approx 0$ more massive galaxies have a lower dust-to-stellar mass ratio (DtS) relative to lower mass galaxies, since dust destruction processes dominate in the high mass, early-type population \citep{DeVis2017HerschelAtlasScalingRelations}.
More importantly, the dust-to-gas ratio (DtG) is found to correlate strongly with gas-phase metallicity \citep[traced by oxygen abundance, e. g.,][]{Remy-Ruyer2015DustEmissionLowMetal}.
This relation highlights the role of ISM dust growth, since more metals drive increased metal accretion into grains. 
Moreover, theoretical models predict that the relation becomes steeper at metallicities larger than a  ``critical metallicity'' $Z_\mathrm{crit}$,
at which the ISM growth rate surpasses the production rate from stellar sources \citep[e. g.,][]{Inoue2011OriginOfDust, Asano2013DustFormationHistory}. 
These models also have an explicit dependence on density in the equations for ISM dust growth, since it's ultimately a high metal density in the vicinity of grains what makes the accretion process efficient \citep[e. g.,][]{Narayanan2026DustGrowthFirstBillionYears}.
However, different models predict different values for $Z_\mathrm{crit}$ \citep[$0.01-1Z_\odot$, see][for a recent review]{Parente2025DustySimReview}, and the observational evidence for a turnover in the relation is still debated \citep[e. g.,][]{DeVis2019DustPediaDtmRatios, Galliano2021NearbyGalaxyDustEvol}.

At redshifts $1<z<3$, the epoch of peak cosmic SFR density or ``cosmic noon'' \citep{MadauAndDickinson2014Review}, our knowledge about the dust budget is significantly more limited.
Samples at $z\approx 2$ with detections of far infrared (FIR) dust emission necessary for measuring \mdust are often biased to the most massive dusty systems \citep[e. g.,][]{Kirkpatrick2017ColdDustSupersample, Donevski2020DustToStellarEvolutionDsfg, Kokorev2021EvolvingIsmStardust}, and thus are not representative of the general galaxy population.
These samples include dusty star-forming galaxies \cite[DSFGs; e. g.,][]{Casey2014DSFGReview}, which have infrared luminosities in excess of \SI{1e12}{\lumsol} and are typically selected in submillimeter wavelengths.
Moreover, the ultraviolet to optical spectrum of DSFGs is highly attenuated \citep[e. g.,][]{Dudzeviciute2020As2udsPropertiesOfSmgs, McKinney2025ScubaDive1}, complicating the accurate determination of stellar masses and metallicities \cite[e. g.,][]{Hainline2011StellarMassSmgs, Simpson2014AlmaCdfsSmgs, Casey2017MosfireDsfgSurvey}.

A more representative account of dust mass at $z\approx 2$ requires targeting much fainter galaxies, those with lower mass and lower metallicity than DSFGs.
Several efforts in this direction have been conducted in the last decades,  especially after the Atacama Large Millimeter/Submillimeter Array (ALMA) became available. 
These campaigns follow a range of different strategies.
For example, some works target gravitationally lensed galaxies, exploiting the magnified flux to probe intrinsically faint galaxies \citep[e.
g.,][]{Saintonge2013MolGasLensedGalaxies, DessaugesZavadsky2015MolGasLensedSFGs, Solimano2021MolecularGasIntermediateMass, Catan2024MolecularBudgetCosmicNoon}.
While these studies have yielded a handful of detections, the effectiveness of this approach is hampered by the rarity of strong-lensing systems \citep[e. g.,][]{Lemon2024StrongLensingSearchReview} and the additional uncertainties introduced by lens modeling \citep[e. g.,][]{Raney2020FrontierFieldsLensingUncertainties, Natarajan2024StrongLensingClustersReview}.
Other studies use the technique of stacking submillimeter data, which has produced a few detections in stellar mass bins with $M_\mathrm{stars}\lesssim\SI{1e10}{\msun}$, but sacrificing knowledge about individual sources, or without any handle on metallicity \citep[e. g.,][]{Santini2014EvolutionDustGasContentGalaxies, Wang2022A3CosmosCensusMolGas, Shivaei2022InfraredSedSubsolar,Jolly2025AlcsDustMassStacks, Hill2026EuclidAverageIRprops, Casey2026DustAverageGalaxyCosmos}.
Another approach is to blindly select faint dust continuum sources from very deep ALMA integrations, as done with the ALMA Spectroscopic Survey in the Hubble Ultra Deep Field \citep[ASPECS; e. g.,][]{Walter2016AspecsPilot, GonzalezLopez2020AspecsDeep1p2mmCounts, Aravena2020AspecsNatureOfFaintDusty}.
However, the narrow field of view of ALMA makes this strategy inefficient.

In this work, we present dust mass measurements of a targeted sample of 25 unlensed star-forming galaxies at $z\approx 2.3$ with robust stellar masses and metallicities as part of the ALMA Chemical Evolution (ACE) Large Program (Shivaei et al. in prep.).
This program was designed to probe $^{12}\mathrm{CO}(3-2)$ molecular line and cold dust continuum emission in galaxies with lower stellar masses (down to \SI{1e9}{\msun}) and lower metallicities (down to $\sim0.3Z_\odot$) than previous samples at this redshift, therefore expanding the parameter space into more typical populations in a systematic way. 
The description and results of the CO observations are presented  in other papers of the ACE series:
\citet{Langan2026AceGfmr} analyzes the molecular gas masses alongside stellar masses as predictors of the gas phase metallicity in the context of gas-regulator models. \citet{Popping2026AceGasScalingRelations}  presents molecular gas scaling relations and average ISM properties of the sample,  while \citet{Geesink2026AceGasToDustRatio}  studies the correlation between continuum and CO emission, and the dust-to-gas mass ratio as a function of metallicity.

The paper is organized as follows: in \autoref{sec:obs} we describe the sample selection, the ALMA observations, and ancillary data.
In \autoref{sec:analysis} we explain how we derive dust masses from the observed continuum flux density and show the resulting dust scaling relations at $z\approx 2$.
In \autoref{sec:discussion} we discuss our findings in the context of the literature across a broad range of redshifts.
We summarize our conclusions in \autoref{sec:conclusions}.
Throughout the paper, we assume a flat cosmology described by $H_0=\SI{67.66}{\kilo\meter\per\second\per\mega\parsec}$, and $\Omega_{m,0}=0.311$ \citep{PlanckCollab2020Parameters}.
Stellar masses and star formation rates assume the \citet{Chabrier2003IMF} initial mass function (IMF).
We define metallicity as the nebular oxygen abundance, $\logoh$, and adopt $\logoh=8.69$ as the solar metallicity \citep{Asplund2009CompositionSun}.

\section{Observations and data}\label{sec:obs}
\subsection{Sample selection and properties}\label{sec:obs:sample}
The ACE sample was selected from the Multi-Object Spectrometer For Infra-Red Exploration (MOSFIRE) Deep Evolution Field survey \citep[MOSDEF;][]{Kriek2015MosdefSurvey}, a Keck/MOSFIRE spectroscopic survey of $\sim1500$ optically selected galaxies at $z=1.4-3.8$.
From the parent sample, galaxies were further selected to (1) be in the Cosmic Evolution Survey (COSMOS) field;  (2) have a spectroscopic redshift between 2 and 2.5; (3) have a S/N$>3$ in the in \halpha, \hbeta, \singlet{O}{iii}{5008}, and \singlet{N}{ii}{6585} lines to enable robust metallicity and SFR estimates; and (4), show no evidence of an active galactic nucleus (AGN) based in X-ray emission, \textit{Spitzer}/IRAC colors, and optical line ratios \citep{Coil2015MosdefAgnDiagnostics, Azadi2017MosdefAgnMultiwavelength, Azadi2018MosdefMidIrExcess, Leung2019MosdefAgnOutflows}.

This resulted in 27 galaxies that were targeted with ALMA Band 6 to detect the dust continuum in program \#2019.1.01142.S (PI: I.
Shivaei, \citealt{Shivaei2022InfraredSedSubsolar}).
Two objects are further removed due to a disjoint dust/optical morphology or an uncertain metallicity, leaving a final sample of 25 galaxies.
Further details on the sample selection are given in the ACE survey overview paper \citep{Shivaei2026AceOverview}.

While metallicities for the ACE sample had already been published \citep{Sanders2015MosdefMassMetallicity, Shivaei2022InfraredSedSubsolar}, here we recomputed them using the \citet{Sanders2026AuroraMetallicity} calibration, that is based on a sample of 139 galaxies at $z\gtrsim2$ with James Webb Space Telescope (JWST)  measurements of direct-method metallicities.
Previous calibrations were based on $z\sim0$ high-redshift analogs rather than actual high-redshift galaxies \citep[e. g.,][]{Bian2018LocalAnalogMetallicityCalib}.
The details of our method to derive metallicities are presented in \citet{Shivaei2026AceOverview}.
Under the new calibration, the ACE galaxies have oxygen abundances ranging from 8.19 (32\% solar) to 8.62 (83\% solar) with an average of 8.40 (51\% solar).

Since ACE galaxies lie in the widely studied COSMOS field, they benefit from a rich photometric dataset of more than 30 broad and medium band filters spanning from rest-frame ultraviolet (UV) to radio wavelengths.
In Shivaei et al. (in prep.) we used photometry from the COSMOS2025 catalog \citep{Shuntov2025CosmosWeb25Catalog}, paired with archival Spitzer and Herschel data, plus the new ALMA photometry. 
Notably, the COSMOS2025 catalog includes JWST data from the COSMOS-Web survey \citep{Casey2023CosmosWebOverview} covering rest-frame optical and near-infrared wavelengths, which are crucial for an accurate estimation of stellar mass.
We then modeled and fitted the spectral energy distribution (SED) with the \textsc{prospector} code \citep{Johnson2021Prospector}, as described in detail in \citet{Shivaei2026AceOverview}.
Briefly, our SED fits include models of both stellar and nebular emission, with a parametric star formation history (delayed-$\tau$), a flexible attenuation curve, and emission from dust according to the \citet{DraineAndLi2007DustModel} model. While the code outputs posterior probability distributions for several parameters, for this paper we were only interested in the stellar mass.
We found stellar masses ranging from \SI{1e9}{\msun} to \SI{4e10}{\msun}, with an average of \SI{1.3e10}{\msun}.

We used the SFRs derived from attenuation-corrected \halpha luminosity reported in \citet{Shivaei2022InfraredSedSubsolar}, where the attenuation was estimated from the Balmer decrement \halpha/\hbeta.
While multiple SFR indicators are available for our sample, including the inferred SFRs from \textsc{prospector}'s output, and the integrated UV+IR luminosity, we preferred \halpha because it is independent of the ALMA data we used for the determination of dust masses. We defer the reader to \citet{Shivaei2026AceOverview} for a detailed comparison between SFR indicators in the ACE survey.
In the context of its parent sample, ACE galaxies lie on and above the star-forming main sequence (MS) derived for the 128 MOSDEF galaxies at $2.09 < z < 2.61$ \citep{Shivaei2015MosdefMainSequence}.
Following \citet{Shivaei2026AceOverview}, the MS prescription applicable to ACE is $\log(\mathrm{SFR}_{\mathrm{H}\alpha}/\si{\msun\per\year}) = 0.58 \times \log(M_\mathrm{stars}/\si{\msun}) - 4.49$, with an observed scatter of 0.36 dex.
With respect to this MS, ACE galaxies show SFR offsets between $-0.1$ and $1.1$ dex.
The high SFRs are likely a consequence of the S/N required for the emission lines in our sample selection.

\subsection{ALMA continuum}\label{sec:obs:alma}
Within the ACE program (\#2024.1.00534.L, PI: I. Shivaei, co-PI: G. Popping), the sample was targeted with ALMA Band 7 (\SI{873}{\micro\meter}) and Band 3 (\SI{3}{\milli\meter}) to obtain unresolved measurements of dust continuum and CO(3-2) line emission, respectively.
The archival Band 6 (\SI{1.3}{\milli\meter}) data was re-reduced together with the new Band 7 data in a consistent way, using the same pipeline version and scripts.
The Band 7 observations reached an average rms of \SI{0.04}{\milli\jansky\per\beam}, while Band 6 averaged \SI{0.03}{\milli\jansky\per\beam}.
This resulted in 17 galaxies detected in Band 7, a considerable increase compared to Band 6, where only four galaxies were detected at S/N$>3$ \citep{Shivaei2022InfraredSedSubsolar}.
In Band 3, the line-masked continuum maps produced no detections.

Briefly, integrated continuum fluxes were extracted from the peak emission in tapered images. The amount of tapering (either \ang{;;2} or \ang{;;3}) was determined for each image via curve-of-growth analysis to ensure the galaxies became unresolved and all the emission was recovered. The rms was then obtained directly from the pixel value distribution of the tapered images after masking all sources in the field of view. We took the rms as the flux error without including the systematic flux calibration uncertainty. More details about the observations and the data reduction are provided in Popescu et al. (in prep.).

Given the significant fraction of non-detections, we performed image-plane stacking in metallicity and stellar mass bins. Details about the stacking are given in \citet{Geesink2026AceGasToDustRatio}  and in \autoref{sec:ap:stacks}.

\subsection{ALMA CO(3-2)}\label{sec:obs:almaco}
The Band 3 datacubes were imaged with natural weighting and a channel width of $\Delta\nu=\SI{7.8125}{\mega\hertz}$ ($\sim\SI{23}{\kilo\meter\per\second}$). The strategy for flux extraction was similar to that used for the continuum observations, where CO(3-2) integrated fluxes were extracted from spatially unresolved data products. Using prior redshift and line width from MOSFIRE's \halpha, we created moment-0 images and extracted the spectrum from the peak if the line was detected or otherwise at the prior optical coordinates of the source. When the line was detected, the frequency integration range and extraction position were adjusted iteratively to maximize the S/N. In these cases, we fitted a Gaussian profile to the line spectrum to measure its integrated flux. Otherwise, upper limits for non-detections were obtained from the spectrum of a beam-sized aperture at the optical coordinates of the source . This procedure resulted in $\sim65\%$ of the sample detected at $>3\sigma$. Further details on the ACE CO(3-2) data reduction and analysis are described in \citet{Langan2026AceGfmr}. 

\section{Analysis}\label{sec:analysis}

\subsection{Continuum flux to dust mass}\label{sec:analysis:dustmass}

Dust mass is dominated by the largest grains from the cold dust population, which reach thermal equilibrium after being heated by the ambient radiation field.
This results in thermal emission in the FIR regime, and its intensity as a function of wavelength can be modeled by a modified blackbody (MBB) curve \citep{Galliano2018IsmDustNearbyReview}.
While different grain populations attain different equilibrium temperatures across the ISM of a galaxy, the galaxy-integrated emission is often well represented by a single-temperature MBB, where the temperature $T_\mathrm{dust}$ is a mass-weighted average of the temperature distribution \citep[e. g.,][]{Magrini2011HeViCSdustToGas, Bianchi2013VindicatingMbb, Nersesian2019DustPediaOldAndYoung, SommovigoAndAlgera2025MultiTemperatureDust}.

Under the MBB model, the total dust mass ($M_\mathrm{dust}$) is related to the observed flux density $S_\nu$ via the equation
\begin{equation}
	\label{eq:dmass_thick}
	M_\mathrm{dust} = \frac{(S_\nu / f_\mathrm{CMB}) D_L^2(z)}{B_\nu(T_\mathrm{dust})(1+z)\kappa_\nu(\beta)}\left(1 - \frac{e^{-\tau(\nu)}}{\tau(\nu)}\right),
\end{equation}
where $f_\mathrm{CMB}$ is the correction factor for the cosmic microwave background (CMB) at redshift $z$ \citep{daCunha2013CmbEffect}, $D_L(z)$ is the luminosity distance to redshift $z$, $B_\nu(T_\mathrm{dust})$ is the Planck function for temperature $T_\mathrm{dust}$, $\kappa_\nu=\kappa_0\left(\nu/\nu_0\right)^\beta$ is the dust grain absorption cross section per unit mass at rest-frame frequency $\nu$, where $\kappa_0$ is the normalization at $\nu_0$ and $\beta$ is the dust emissivity index.
The optical depth $\tau(\nu)=(\nu/\nu_\mathrm{thick})^\beta=(\lambda_\mathrm{thick}/\lambda)^\beta$ scales in the same way as $\kappa_\nu$, and it is normalized at the rest wavelength $\lambda_\mathrm{thick}$ where it reaches unity. The dust mass surface density is related to optical depth via $\tau_\nu = \kappa_\nu \Sigma_\mathrm{dust}$.
Optical depth mostly affects the blue side of the MBB spectrum, and therefore $\lambda_\mathrm{thick}$ is highly degenerate with $T_\mathrm{dust}$. In dusty galaxies with good FIR coverage, $\lambda_\mathrm{thick}$ averages $\sim\SI{75}{\micro\meter}$ \citep{Simpson2017MultiwavePropsA2Uds}, with values as high as $\sim\SI{240}{\micro\meter}$ in the most extreme and compact dusty starbursts \citep[e. g.,][]{Rangwala2011Arp220, Cortzen2020ColdDustGn20}.
Here, we are probing the continuum at $\lambda_\mathrm{rest} \gtrsim \SI{250}{\micro\meter}$ therefore the effect of optical depth is limited. For example, an MBB model with $\lambda_\mathrm{thick}=\SI{100}{\micro\meter}$ yields a $M_\mathrm{dust}$ that is only 12\% higher than an optically thin MBB at a fixed temperature (\SI{25}{\kelvin}), $\beta$, and  \SI{250}{\micro\meter} flux density. Since we do not expect the dust optical depth in our sample to be higher than that of DSFGs\footnote{Within our adopted $\kappa_\nu$ model, a $\lambda_\mathrm{thick}=\SI{100}{\micro\meter}$ translates to $\Sigma_\mathrm{dust}=\SI{1.8e8}{\msun\per\kilo\parsec\squared}$. To attain such a high density, the most compact galaxy in our sample (ID-3626 with effective radius of $\sim\SI{0.7}{\kilo\parsec}$, \citealt{Price2016MosdefHalphaKinematicsStructures}) should have $M_\mathrm{dust}\sim\SI{6e8}{\msun}$, but its non detection in Band 7 implies $M_\mathrm{dust}<\SI{6.5e7}{\msun}$.}, we decided to work in the optically thin limit, where \autoref{eq:dmass_thick}  becomes
\begin{equation}
	\label{eq:dmass}
	M_\mathrm{dust} = \frac{(S_\nu / f_\mathrm{CMB}) D_L^2(z)}{B_\nu(T_\mathrm{dust})(1+z)\kappa_\nu(\beta)}.
\end{equation}


Most of the sample is detected only in Band 7, thus we refrained from fitting MBBs with $T_\mathrm{dust}$ or $\beta$ as free parameters.
Instead, we obtained a single-band $M_\mathrm{dust}$ estimate from \autoref{eq:dmass} by fixing $T_\mathrm{dust}=\SI{25}{\kelvin}$ and $\beta=2.08$ (see \autoref{tab:mdust}).
These values were informed by the analysis of the average SED of the ACE sample presented in Popescu et al. (in prep.).
Their work includes stacked constraints on Band 3, Band 6, and Band 7 which suggested $\beta \approx 2$, contrary to the previous notion that $\beta<2$ was more appropriate for this subsolar metallicity sample \citep{Shivaei2022InfraredSedSubsolar}.
With this in mind, we followed the recommendation of \citet{Bianchi2013VindicatingMbb} and chose a set of $\beta$, $\kappa_0$, and $\lambda_0$ that are compatible with a detailed dust grain mixture model.
In particular, we adopted the combination of  $\beta=2.08$ and  $\kappa_0 = \SI{0.4}{\meter\squared\per\kilogram}$ at $\lambda_0=\SI{250}{\micro\meter}$ ($\nu_0=\SI{1199}{\giga\hertz}$) , a pair of values that come from the \citet{Draine2003InterstellarDustGrainsReview} model of Milky Way's dust.
The underlying assumption is that the optical properties of the dust particles in the ACE galaxies are the same as those of the Milky Way. The validity of this assumption is discussed in \autoref{sec:discussion:caveats}.

The dust temperature was set to \SI{25}{\kelvin} as suggested by \citet{Scoville2016MainSequenceAndGasMass} and adopted in other studies of massive, metal-rich galaxies \citep[e . g.,][]{Kirkpatrick2017ColdDustSupersample}. We acknowledge, however, that local low-metallicity dwarfs tend to have higher dust temperatures and broader SEDs \citep[e. g.,][]{Remy-Ruyer2015DustEmissionLowMetal}. In fact, MBB fits to composite SEDs of the ACE sample in metallicity bins yield temperatures between \SI{28}{\kelvin} and \SI{42}{\kelvin} with a clear anticorrelation with metallicity (Popescu et al. in prep.).
The key point here is that these fitted temperatures are inevitably biased to the most active regions within galaxies (i. e., luminosity-weighted temperatures), and therefore may not represent the temperature of the bulk of the dust mass.
Still, small variations in the dust temperature have a sizable effect in the continuum emission at $\lambda_\mathrm{rest}=\SI{250}{\micro\meter}$: 
at an observed wavelength of \SI{873}{\micro\meter} and $z=2.3$, a \SI{5}{\kelvin} colder (hotter) $T_\mathrm{dust}$ results in a $\sim50\%$ higher (lower) dust mass for a given flux density.

In \autoref{sec:ap:tdz} we explored the effect of assuming that the mass-weighted $T_\mathrm{dust}$ has a similar variation with metallicity as the luminosity-weighted $T_\mathrm{dust}$ and compare it with our fiducial choice of $T_\mathrm{dust}=\SI{25}{\kelvin}$.

Besides $T_\mathrm{dust}$, $\kappa_0$ is an even larger source of systematic uncertainty \citep[see discussions in][]{Shivaei2022InfraredSedSubsolar, Casey2026DustAverageGalaxyCosmos}.
Dust opacities at $\lambda_\mathrm{rest}=\SI{250}{\micro\meter}$ vary up to a factor of three among the different dust mixture models available in the literature \citep[e. g.,][]{DraineAndLi2007DustModel, Galliano2011DustModelLMC, Jones2017ThemisDustModel, Guilet2018DustModelsPlanck, HensleyAndDraine2023AstrodustPlusPah}.

We computed \mdust from Band 7 fluxes ($S_{\SI{873}{\micro\meter}}$) for the 17 detected galaxies using \autoref{eq:dmass}.
For ID-8515 we report the dust mass based on the Band 6 detection, as it was not detected in Band 7.
The errors were computed by randomly sampling from two independent normal distributions: one representing the measured flux density and its uncertainty, and the other accounting for variations in the dust temperature, with a mean value of \SI{25}{\kelvin} and standard deviation of \SI{5}{\kelvin} as in \citet{Shivaei2022InfraredSedSubsolar} and \citet{Popping2023DustToGasVsMetallicity}.
For the undetected sources, we used the ALMA Band that gives the lowest upper limit on the mass given our aforementioned assumptions.
The results are given in \autoref{tab:mdust}, together with the redshift, stellar mass, SFR, and oxygen abundance.
We find that dust masses for the 17 detected galaxies range from \SI{5.4e7}{\msun} to \SI{2.8e8}{\msun}, with a median value of \SI{6.4e7}{\msun}.
The remaining sources have a median upper limit of $M_\mathrm{dust} < \SI{5.8e7}{\msun}$.

\subsection{Correlation analysis and linear regression}\label{sec:analysis:scaling}

To quantify the strength of any relations between the quantities of interest: $M_\mathrm{dust}$, $M_\mathrm{star}$, SFR, \logoh, etc. (\autoref{tab:regression} columns 1 and 2), we computed Kendall's $\tau$ partial rank correlation coefficient and its corresponding two-sided $p$-value following \citet{AkritasAndSiebert1996PartialKendallTauCensored} and \citet{Flury2022LozLyCSurvey2}.
The advantage of this method over other correlation diagnostics is that it can be applied to datasets with left-censoring (i. e., upper limits).
Next, we fitted powerlaw scaling relations to our data (again including upper limits on $M_\mathrm{dust}$) by running a linear regression with the  \verb|PyMC| Bayesian framework \citep{Patil2010PyMC}, following the same procedure as \citet{Langan2026AceGfmr}.
For each pair of quantities $x$ and $y$, we modeled $\log(y)$ as
\begin{equation}\label{eq:linear}
	\log(y) = a \times \log\left(\frac{x}{x_0}\right) + b + \sigma_\mathrm{int},
\end{equation}
where $x_0$ is the mean value of the observed $x$.
In addition to the coefficients $a$ and $b$, we modeled the data to be normally distributed around this relation with an intrinsic scatter $\sigma_\mathrm{int}$.
We adopted a Gaussian likelihood function that takes into account normally distributed measurement errors in both $\log(x)$ and $\log(y)$.
For upper limits of $\log(y)$ we assumed a cumulative Gaussian likelihood corresponding to the probability that the true value lies below the detection threshold.
Then we adopted the following weakly informative priors:
\begin{align*}
	&a \sim \mathrm{Normal}(\mu=0, \sigma=2) \\
	&b \sim \mathrm{Normal}(\mu=\left<\log(y)\right>, \sigma=2) \\
	&\sigma_\mathrm{int} \sim \mathrm{HalfNormal}\footnotemark(\sigma=2). \\
\end{align*}\footnotetext{A half-normal distribution is a normal distribution centered at zero with its support restricted to positive values (i. e., from zero to infinity).
}
Finally, we sampled the posterior distribution using the No U-Turn Sampler (NUTS) Markov Chain Monte Carlo (MCMC) algorithm as implemented within \verb|PyMC|.
We ran the sampler with eight chains, 1000 tuning steps, and 400 draws per chain.
The results for six different pairs of $x$ and $y$ are listed in \autoref{tab:regression}.

\begingroup
\renewcommand*{\arraystretch}{1.5}
\begin{table*}[!hbt]
  \centering
  \caption{Results of the correlation test and linear regression described in \autoref{sec:analysis:scaling}.}
  \label{tab:regression}
  \begin{tabular}{cccccccc}
  \hline\hline
  $x$ & $y$ & $\tau$ & $p$ &$\log(x_0)$ & $a$ & $b$ & $\sigma_\mathrm{int}$ \\
  \hline
  $M_\mathrm{stars}$ & $M_\mathrm{dust}$ & 0.32 & 0.023 & 9.98 & $0.31_{-0.22}^{+0.23}$ & $7.88_{-0.08}^{+0.07}$ & $0.30_{-0.07}^{+0.09}$\\
$12+\log(\mathrm{O}/\mathrm{H})$ & $M_\mathrm{dust}$ & 0.45 & 0.002 & 8.40 & $1.90_{-0.47}^{+0.51}$ & $7.89_{-0.06}^{+0.06}$ & $0.17_{-0.06}^{+0.08}$\\
$\mathrm{SFR}$ & $M_\mathrm{dust}$ & 0.50 & \num{5.0e-04} & 1.79 & $1.03_{-0.29}^{+0.33}$ & $7.86_{-0.06}^{+0.07}$ & $0.17_{-0.07}^{+0.08}$\\
$12+\log(\mathrm{O}/\mathrm{H})$ & $M_\mathrm{dust}/M_\mathrm{stars}$ & 0.17 & 0.243 & 8.40 & $1.03_{-0.80}^{+0.79}$ & $-2.12_{-0.10}^{+0.10}$ & $0.40_{-0.09}^{+0.12}$\\
$12+\log(\mathrm{O}/\mathrm{H})$ & $M_\mathrm{dust}/\mathrm{SFR}$ & 0.43 & 0.003 & 8.40 & $1.56_{-0.49}^{+0.53}$ & $6.05_{-0.06}^{+0.06}$ & $0.11_{-0.05}^{+0.07}$\\
$\mathrm{sSFR}$ & $M_\mathrm{dust}/M_\mathrm{stars}$ & 0.57 & \num{7.2e-05} & -8.19 & $0.73_{-0.13}^{+0.15}$ & $-2.12_{-0.05}^{+0.06}$ & $0.10_{-0.07}^{+0.09}$

  \\ \hline
  \end{tabular}
  \tablefoot{The values shown for $a$, $b$, and $\sigma_\mathrm{int}$ correspond to the median of the posterior probability function along with their respective 68\% confidence intervals. All masses are in units of \si{\msun}, and SFR has units of \si{\msun\per\year}. Specific SFR, $\mathrm{sSFR}=\mathrm{SFR}/M_\mathrm{stars}$, has units of \si{\per\year}.}
\end{table*}
\endgroup

\subsection{Comparison samples}\label{sec:analysis:comparison}

Here we describe our compilation of literature data at $z=0$,  $z\sim2$, and beyond.
Whenever possible, we used datasets with $M_\mathrm{stars}$, $M_\mathrm{dust}$, SFR, and \logoh~obtained with consistent methodologies and similar assumptions as the ones we used for ACE. Stellar masses and SFRs were converted to the \citet{Chabrier2003IMF} IMF according to the conversion factors of \citet{MadauAndDickinson2014Review} when appropriate.

The main sample at $z=0$ was taken from \citet{Galliano2021NearbyGalaxyDustEvol}, consisting of 764 galaxies from the DustPedia sample \citep{Clark2018DustPediaPhotometry} plus 34 additional galaxies from \textit{Herschel}'s Dwarf Galaxy Survey \citep[DGS;][]{Madden2013DwarfGalaxySurvey}.
Instead of recomputing the dust masses, we adopt the published values which have been obtained through fitting of the \textsc{THEMIS} model \citep{Jones2017ThemisDustModel} to all the available IR bands. We only apply a small correction factor of $0.64/0.4$ to $M_\mathrm{dust}$ since the THEMIS model assumes an opacity of $\kappa_0 = \SI{0.64}{\square\meter\per\kilo\gram}$ at \SI{250}{\micro\meter}. 

In order to populate the high SFR regime, we complemented the $z=0$ sample with $\sim200$ luminous infrared galaxies (LIRGs, $L_\mathrm{IR}>\SI{1e11}{\lumsol}$) from the Great Observatories All-sky LIRG Survey \citep[GOALS;][]{Armus2009Goals}.
Here, we computed the dust masses by scaling our fiducial MBB to the flux recorded by \textit{Herschel}'s Spectral and Photometric Imaging Receiver (SPIRE) \SI{350}{\micro\meter} band, from photometry published in \citet{Chu2017GoalsHerschelPhotometry}.
The stellar masses were obtained via SED fitting and provided in a private communication based on the results of \citet{DiazSantos2010SpatialExtentUlirgs, DiazSantos2013GoalsCiiDeficit} and \citet{Howell2010GoalsComparisonUvAndFir}.
We computed SFRs multiplying the total IR luminosity by \SI{1.4e-10}{\msun\per\year\per\lumsol} \citep{Murphy2011DustObscuredSfr, Schaerer2020AlpineCiiSfrRelation} and by the $(1-f_\mathrm{AGN})$ factor, where $f_\mathrm{AGN}$ is the bolometric AGN fraction \citep{DiazSantos2017GoalsFirLineSurvey}.

Beyond $z=0$, we included data from the ``Supersample'' of 51 LIRGs spanning $0.5<z<3$ from \citet{Kirkpatrick2017ColdDustSupersample}. These galaxies have well sampled FIR SEDs plus  mid-IR spectra from \textit{Spitzer} that ensures a proper separation between AGN and star-forming contributions to the IR luminosity. The authors obtain independent SFR and $M_\mathrm{dust}$ estimates by using their longest wavelength datapoint ($\lambda_\mathrm{rest} >\SI{250}{\micro\meter}$) exclusively to measure $M_\mathrm{dust}$. They assume an optically thin MBB with fixed $T_\mathrm{dust}=\SI{25}{\kelvin}$, but with a different $\kappa_\nu$ model. Therefore we scaled the published masses to match our adopted model. 

We complemented these data with a sample of DSFGs from the SCUBADIVE project \citep{McKinney2025ScubaDive1}.
These galaxies are bright ($S_{\SI{850}{\micro\meter}}>\SI{2}{\milli\jansky}$)  submillimeter galaxies (SMGs) in the COSMOS field \citep{Simpson2019CosmosScuba2} that now have robust stellar masses thanks to JWST imaging and careful SED fitting.
We further refined the sample by selecting  all DSFGs with spectroscopic redshifts and $2<z<3$, and available fluxes from archival ALMA Band 6 or Band 7 data.
The latter were used for computing the dust mass using the MBB scaling mentioned above.
From this sample of 24 DSFGs, we further excluded the two sources hosting known X-ray AGN by crossmatching to the C-COSMOS catalog \citep{Elvis2009ChandraCosmosOverview}. 
Metallicities were not available for neither GOALS, Supersample, nor SCUBADIVE sources, but we expect them to range from near-solar to super-solar \citep[e. g.,][]{Chartab2022FirLinesMetallicityUlirgs}.

To compare with less extreme galaxies at  $z\approx 2$, we also included the continuum-detected sources from the deep ASPECS \SI{1.2}{\milli\meter} map \citep{GonzalezLopez2020AspecsDeep1p2mmCounts} that are not classified as X-ray AGN in \citet{Luo2017ChandraDeepFieldSouthCatalog}. We divided the sample in two redshift bins, $1<z_\mathrm{spec}<2$ (26 sources), and $2<z_\mathrm{spec}<3$ (11 sources).
We collected all the available photometry from \citet{Robertson2026JadesDr5Catalog} and fitted the SED following the same procedure outlined in \citet{Shivaei2026AceOverview}, except we included templates for AGN contribution in the mid IR \citep{Nenkova2008AgnDustyToriClumpyMediaI}.
Due to the lack of \halpha data, the SFRs were obtained from the output of \textsc{prospector} (\SI{10}{\mega\year} averages).
Twelve of our selected ASPECS sources have JWST/NIRSpec grating spectra \citep{Kiyota2026JwstFaintSmgs} from which we measure metallicities following the same method and calibration used with ACE (see \autoref{sec:obs:sample} and \citealt{Shivaei2026AceOverview}).

Finally, we collected measurements of UV-luminous galaxies at higher redshift from the ALPINE-CRISTAL JWST \citep[$z\approx 5$, ][]{Faisst2026AlpineCristalJwstOverview} and the REBELS-IFU \citep[$z\approx7$,][]{Algera2026RebelsIfuDustBuildup} surveys.
For the 19 ALPINE-CRISTAL sources, we computed dust masses by scaling our fiducial MBB model to the observed rest-frame \SI{158}{\micro\meter} continuum flux \citep{Mitsuhashi2024AlmaCristalDustSizes}, while the metallicities were recomputed in the \citet{Sanders2026AuroraMetallicity} calibration using published strong line ratios \citep{Faisst2026AlpineCristalJwstOverview}. Stellar masses and dust-corrected \halpha SFRs were also taken from \citet{Faisst2026AlpineCristalJwstOverview}.
For the 12 REBELS-IFU sources in \citet{Algera2026RebelsIfuDustBuildup}, we kept their reported $M_\mathrm{dust}$ unchanged. This is because (1)  $T_\mathrm{dust}=\SI{25}{\kelvin}$ is no longer a reasonable assumption at $z\approx7$ given that $T_\mathrm{CMB}(z=7)=\SI{21.8}{\kelvin}$, so they assume $T_\mathrm{dust}=\SI{45}{\kelvin}$ except for the two sources where they can constrain $T_\mathrm{dust}$ and $\beta$ from the data; and (2), the $\kappa_0$ they adopt comes from the same dust mixture model, so no correction was needed. In turn, metallicities were not recomputed since the \citet{Sanders2026AuroraMetallicity} calibration is poorly constrained at $z\approx7$. Stellar masses were also taken at face value from \citet{Algera2026RebelsIfuDustBuildup} while the SFRs are taken from  dust-corrected \halpha when both \halpha and \hbeta are observed, otherwise we took the lower limit inferred from the \hbeta line alone \citep{Rowland2026RebelsIfuMetal}.

\section{Results and discussion}\label{sec:discussion}

\begin{figure*}[!htb]
	\resizebox{\hsize}{!}
	{\includegraphics {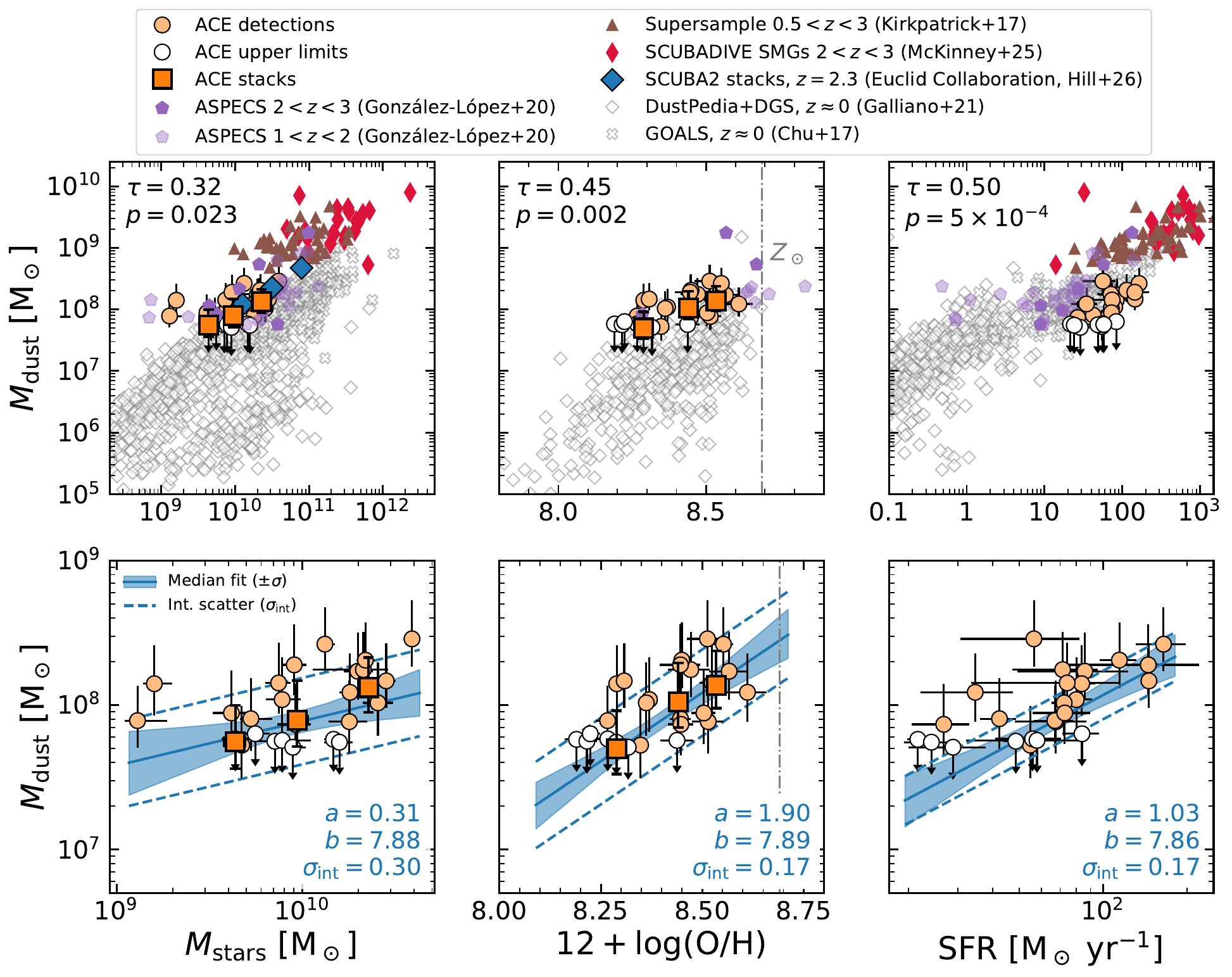}}
	\caption{
		Dust masses of ACE galaxies as a function of stellar mass (\textit{left}), metallicity (\textit{center}), and SFR (\textit{right}).
		In the left and central panels, the bright orange squares indicate the stacked ACE dust masses in bins of stellar mass and metallicity, respectively (see \autoref{sec:ap:stacks}) .
		The top row  shows the ACE data in comparison to a compilation of $z\approx0-3$ galaxies from the literature \citep{Kirkpatrick2017ColdDustSupersample, Chu2017GoalsHerschelPhotometry, Galliano2021NearbyGalaxyDustEvol,  GonzalezLopez2020AspecsDeep1p2mmCounts, McKinney2025ScubaDive1, Hill2026EuclidAverageIRprops}, while the bottom row shows a zoom-in view of the ACE data alone. 
		The errorbars of the literature data are omitted for visual clarity.
		In the upper left corner of each top row panel, we display Kendall's $\tau$ correlation coefficient and its $p$-value (\autoref{sec:analysis:scaling}).
		In the bottom panels, the solid blue line indicates the median posterior fit powerlaw to the ACE data (see \autoref{sec:analysis:scaling}), with parameters listed in the legend in the lower right corner. The blue area covers the 68\% posterior confidence interval, while the dashed blue lines illustrate the intrinsic scatter around the median fit line.  In the central panels, the vertical dash-dotted line marks the solar metallicity.
	}
	\label{fig:mdustmulti}
\end{figure*}

\subsection{Dust mass at $z\approx 2.3$}\label{sec:discussion:dmass}
In \autoref{fig:mdustmulti} we show the dust mass as a function of stellar mass, gas-phase metallicity, and SFR.
In all three cases we see a mild upward trend with each of these quantities, despite the limited dynamic ranges spanned by our data.
When compared to the low-redshift compilation sample, ACE galaxies occupy a distinct locus in the three parameter spaces.
In the leftmost top panel, at a given stellar mass, ACE galaxies have four to nine times larger dust masses on average than low-redshift galaxies, although there is overlap with the upper range of the dust mass distribution.
Our fiducial dust masses are consistent with the stacked measurements from \citet{Hill2026EuclidAverageIRprops} at $z=2.3$ and the with the low mass end of the ASPECS survey, but remain significantly below massive DSFGs (the Supersample, with a median $M_\mathrm{dust}=\SI{3.8e8}{\msun}$; and SCUBADIVE, with a median  $M_\mathrm{dust}=\SI{1.9e9}{\msun}$) that can reach up to almost \SI{1e10}{\msun}.

For the ACE sample alone, we measure a modest Kendall's $\tau$ of $0.32$, but  with a $p=0.023$ lower than the typical threshold of $0.05$, hence  we can plausibly reject the null hypothesis\footnote{Here, the null hypothesis states that the quantities $x$ and $y$ are uncorrelated ($\tau=0$) in the underlying population.}. However, our regression analysis finds a sublinear slope of $0.3\pm0.2$ in the $M_\mathrm{dust}-M_\mathrm{stars}$ plane (see \autoref{tab:regression}), significantly lower than reported at $z=0$ \citep[$a=0.78$,][]{deLooze2020JingleIVDustScaling}.
This shallow slope is likely driven by the two outliers ID-8515 and ID-3666 at the low stellar mass end.

In the top central panel of \autoref{fig:mdustmulti}, we observe a similar offset for \mdust against metallicity, with ACE galaxies showing three to eight times higher \mdust than local galaxies at a given metallicity.
The correlation tests yields $\tau=0.45$ ($p=0.002$), and we find a superlinear slope of $1.9\pm0.5$ (see \autoref{tab:regression}) that is consistent with the value $2.6\pm0.1$ reported by \citet{Remy-Ruyer2015DustEmissionLowMetal} on a sample of $z=0$ galaxies.

In the right panels, we show dust mass against SFR.
ACE galaxies have higher SFRs than low-redshift counterparts at a fixed stellar  mass, owing partly to the evolution of the MS \citep[e. g.,][]{Popesso2023MainSequenceAcrossCosmicTime}, but also to the fact that our selection criteria picks galaxies with SFRs on and above the MS (see \autoref{sec:obs:sample}).
The SFRs of ACE galaxies are only comparable to those of LIRGs in the local Universe (e. g., galaxies from the GOALS survey), which have larger stellar masses.
At cosmic noon, the SFRs of the ASPECS sample overlap with those of ACE.
Interestingly, at a fixed SFR, local LIRGs and ASPECS galaxies have roughly three and six times the dust mass of ACE galaxies, respectively (see top right panel of \autoref{fig:mdustmulti}).

A possible cause for these offsets is the different SFR tracer used for each sample relative to ACE. 
In GOALS, the SFR is estimated from the total IR luminosity (see \autoref{sec:analysis:comparison}), while in ACE the SFRs come from dust-corrected \halpha luminosities. 
The latter are expected to trace recent star formation (within $<\SI{10}{\mega\year}$), hence if a galaxy experienced a recent burst, its \halpha-based SFR will be higher than indicators probing a longer timescale such as the IR luminosity ($\sim \SI{100}{\mega\year}$).
However, \citet{Shivaei2026AceOverview} find that UV+IR SFRs are consistent with H$\alpha$ SFRs within the ACE sample.
In contrast, they  also find that the \SI{10}{\mega\year}-averaged SFRs obtained from \textsc{prospector} SED fitting are systematically $2.7$ times lower than H$\alpha$ SFRs.
This could partially alleviate the offset with respect to the ASPECS $2<z<3$ galaxies, although it will remain at a factor of $\approx2-3$.
A more complete explanation involves the fact that unlike ACE, both GOALS and ASPECS are IR-selected samples, therefore they could be biased toward high dust masses.

Despite the possible systematic offsets between different samples, we find a correlation coefficient of $\tau=0.5$ ($p=\num{5e-4}$) within the  ACE data, and our regression analysis favors a remarkably linear relation between $M_\mathrm{dust}$ and SFR with a slope of \num{1.0+-0.3}.
This is consistent with the slope of \num{1.11+-0.01} found by \citet{DaCunha2010DustMassSfrSdssIras} for a large sample of galaxies at $z=0$.
\citet{Hjorth2014DustSfrRelation} interpret this relation as the result of an evolutionary sequence, where dust mass responds proportionately to starburst and quenching processes:
as a galaxy enters a starburst phase, the rise in SFR translates into a rise in the number of supernovae enriching the ISM with dust.
At the same time, the SFR is supported by large amounts of gas that further produce dust via metal accretion.
This is balanced by astration, and sputtering in supernova shocks, two processes that are directly linked to SFR.
Afterwards, the galaxy may enter a quenching phase, and if the SFR decline is due to the ejection of gas in outflows, a proportionate amount of dust mass will also be ejected.
Other quenching channels, such as the disruption of the gas supply, can decrease the SFR but retain the dust, and therefore lead to an horizontal transition in the $M_\mathrm{dust}-\mathrm{SFR}$ diagram, contributing to the scatter of the relation.

Another interpretation relies on $M_\mathrm{dust}$ being a tracer of the molecular gas mass, $M_\mathrm{mol}$.
In this view, the $M_\mathrm{dust}-\mathrm{SFR}$ relation is simply a proxy of the integrated Schmidt-Kennicutt relation \citep{Daddi2010SfrLawsHighZ, Genzel2010MainSequence}, between $M_\mathrm{mol}$ and SFR.
This is supported by the tight relation between dust continuum and CO(3-2) luminosities found in the ACE sample, although converting these luminosities to $M_\mathrm{dust}$ and $M_\mathrm{mol}$ increases the scatter due to calibration uncertainties and metallicity-dependent conversion factors \citep{Geesink2026AceGasToDustRatio}.

\subsection{The effect of metallicity}\label{sec:discussion:nometaltrend}

\begin{figure*}[ht!]
	\resizebox{\hsize}{!}
	{\includegraphics {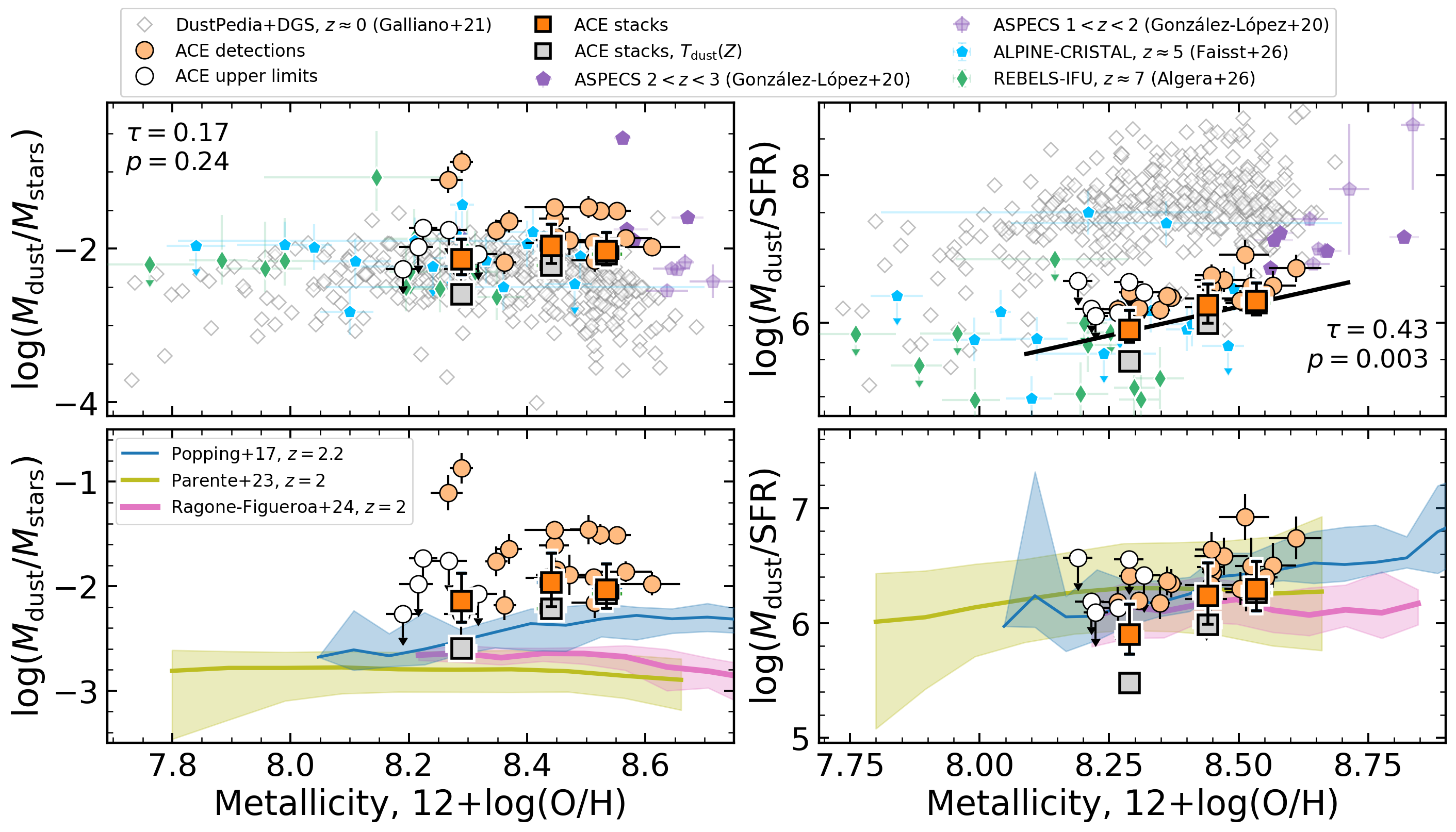}}
	\caption{
		Relative dust content of ACE galaxies versus metallicity.
	\textit{Top left:} Dust-to-stellar mass ratio against metallicity with literature data (blue pentangons: \citealt{Faisst2026AlpineCristalJwstOverview}, green diamonds: \citealt{Algera2026RebelsIfuDustBuildup}).
		Orange and gray markers are the same as in \autoref{fig:mdustmulti}.
		The solid gray squares correspond to our binned measurements with the assumption of a metallicity-dependent $T_\mathrm{dust}$.
		Since we find no significant correlation between DtS and metallicity for our fiducial assumption ($T_\mathrm{dust}=\SI{25}{\kelvin}$), we do not plot the best fit powerlaw from our regression analysis.
		\textit{Top right:} $M_\mathrm{dust}/\mathrm{SFR}$ ratio against metallicity  with literature data.
		The markers and symbols are the same as in the top left panel.
		The best fit powerlaw is shown in a solid black curve.
		\textit{Bottom row:} Same as the corresponding top panels, but the ACE data are compared to cosmological simulations.
		Solid lines indicate the median trend lines from each model in bins of metallicity.
		We display in solid lines the  predictions from the \citet{RagoneFigueroa2024MuppiSim} hydrodynamic simulation, as well as the \citet{Popping2017DustEvolutionSam} and  \citet{Parente2023CosmicDustSemiAnalytic} SAMs.
		The shaded areas around each line correspond to the 16th and 84th percentiles of the simulated galaxy distribution.
	}
	\label{fig:dts_vs_oh}
\end{figure*}

In this section we explore the role of metallicity in explaining our results.
In \autoref{fig:mdustmulti} we saw that ACE galaxies follow a steep trend between metallicity and dust mass.
However, this could be a consequence of the \mdust-\mstar relation and the \mstar-metallicity relation \citep[MZR; e. g.,][]{Tremonti2004MassMetallicityRelation}. To control for the \mstar dependence, we look instead at the DtS ratio.

While the DtS ratio has received comparatively less attention than the dust-to-gas ratio in the literature, stellar mass remains a much more accessible quantity with respect to gas probes that require very expensive observations. Previous work on cosmic noon massive galaxies and DSFGs had limited access to robust metallicity indicators and therefore focused on $M_\mathrm{stars}$, finding an inverse relationship of the form $M_\mathrm{dust}/M_\mathrm{stars} \propto M_\mathrm{stars}^{-\gamma}$ \citep[e. g., ][with $\gamma\lesssim0.6$ ]{Donevski2020DustToStellarEvolutionDsfg}, typically steeper than the one describing low redshift galaxies \citep[e. g.,][with $\gamma=0.22$]{deLooze2020JingleIVDustScaling}. This relation is usually understood as a reflection of the fact that galaxies grow in stellar mass by consuming their ISM, yet the slope is highly dependent on the star formation history and the timescales of ISM replenishment \citep[e. g.,][]{Calura2017DtSevolution}. For reference, our $M_\mathrm{dust}-M_\mathrm{stars}$ fit to the ACE data implies a steep $\gamma=0.7\pm0.2$.

Crucially, stellar mass is connected to the total mass of metals ever formed in a galaxy (assuming a universal IMF and supernovae yields). In an idealized ``closed box'' galaxy without outflows nor inflows of baryons, the metal mass would be directly proportional to the stellar mass, and the DtS ratio will closely trace the dust-to-metal ratio and thus be highly informative of the balance between dust creation and destruction. In a more realistic scenario, both metal and dust content are modulated by inflows, outflows, and other processes, making the interpretation of DtS less straightforward \citep[e. g.,][]{Tan2014DustGasProtocluster, Imara2018ModelDustTemps}.

Despite these complications, the ACE sample gives us the unique opportunity to directly compare DtS against metallicity (top left panel of  \autoref{fig:dts_vs_oh}), unlike most works that can only compare DtS to $M_\mathrm{stars}$ or specific SFR \citep[sSFR=SFR/$M_\mathrm{stars}$; e. g.,][]{Hunt2014HerschelGammaRayHosts, Martis2019DustMassiveGalaxiesUltraVista, Burgarella2020DustConstraintsLbgs}.
The DtS of the ACE sample sits $\approx 0.4$ dex above local galaxies at the same metallicity, and scatters around $\log(M_\mathrm{dust}/M_\mathrm{stars})=-2.0$ (median of the detected sample).
The ACE values are also strikingly similar to those of the $z\approx5$ and $z\approx7$ samples, even when these probe down to much lower metallicities, suggesting little to no evolution of DtS above $z=2$.
Interestingly, neither ACE nor any $z>0$ literature sample show significant trends against metallicity in this parameter space.
In particular, our correlation test to the ACE data yields a large $p$-value (see \autoref{tab:regression}) that prevents us from ruling out the null hypothesis (i. e., no correlation). In other words, the data are consistent with a constant DtS within the range of metallicities probed by ACE.

We note, however, that this behavior heavily relies on the assumed $T_\mathrm{dust}$. Our alternative, metallicity-dependent $T_\mathrm{dust}$ model naturally introduces a positive correlation between DtS and metallicity (illustrated by the solid gray squares, see also \autoref{sec:ap:tdz}). In any case, a flat or slightly positive relation are somewhat unexpected at first glance, considering the inverse relation between DtS and $M_\mathrm{stars}$ together with the MZR. But a flat relation can be achieved if the dust builds up on similar timescales than the stellar mass regardless of metallicity. Instead, a positive relation suggests that dust buildup happens on relatively shorter timescales at high metallicity: a possible mechanism is ISM  grain growth, since the it becomes more efficient at higher metallicity and therefore can contribute to outpace the rate of star formation.  

The logical next step is to look at the  $M_\mathrm{dust}/\mathrm{SFR}$ ratio as a function of metallicity, displayed in the top right panel of \autoref{fig:dts_vs_oh}.
Here, the ACE data follow a stronger trend ($\tau=0.43$, $p=0.003$), with a fitted powerlaw slope of $1.55\pm0.5$ and a scatter of only $\sigma_\mathrm{int}=0.12$ dex (see \autoref{tab:regression}).

With respect to redshift evolution, we can see $\sim 1$ dex drop in \mdust/SFR between $z=0$ \citep{Galliano2021NearbyGalaxyDustEvol} and $z=2$ (this work).
A direct explanation for this offset is the evolution of the MS: at a fixed stellar mass, the SFR increases 1.4 dex from $z=0$ to $z=2$ \citep{Popesso2023MainSequenceAcrossCosmicTime}. Therefore, if DtS evolves weakly within this redshift range (as suggested by the top left panel of  \autoref{fig:dts_vs_oh}), the $M_\mathrm{dust}/\mathrm{SFR}$ ratio will decrease proportionally to the SFR increase.
At higher redshift, the ALPINE-CRISTAL and REBELS samples are widely spread across the diagram, but their average position below ACE also points to the evolving normalization of the MS.

At first glance, since the gas depletion timescale $t_\mathrm{dep}=M_\mathrm{mol}/\mathrm{SFR}$ does not vary significantly as a function of metallicity nor MS offset in the ACE sample \citep{Popping2026AceGasScalingRelations}, the \mdust/SFR becomes  a proxy of the more fundamental $M_\mathrm{dust}/M_\mathrm{mol}$ (DtG) ratio.
In fact, \citet{Geesink2026AceGasToDustRatio} find a positive correlation between DtG and metallicity in the ACE sample with a powerlaw slope of $\approx 1.2$ (cf. their Figure 4) that is consistent with our fiducial $M_\mathrm{dust}/\mathrm{SFR}-Z$ slope. However, while the slopes of both $t_\mathrm{dep}-Z$ and DtG$-Z$ relations  ultimately depend on the assumed CO-to-H$_2$ conversion factor $\alpha_\mathrm{CO}(Z)$ \citep[see discussion in][]{Langan2026AceGfmr}, our observed  $M_\mathrm{dust}/\mathrm{SFR}-Z$ positive correlation also points to a more efficient dust buildup (per unit SFR or $M_\mathrm{mol}$) with increasing metallicity.

Altogether, the DtS and \mdust/SFR results suggest a potential effect of metallicity on the assembly of dust mass--particularly when assuming the metallicity-dependent temperature model. This is consistent with a scenario where dust buildup is dominated by ISM growth, where higher metallicity makes accretion more efficient. Alternatively, the dust buildup is dominated by stellar sources and metallicity only enhances dust survival (e. g., via more efficient cooling) or increases the dust yields. The ISM growth contribution in such a scenario should nevertheless be significant,  because chemical models that do not include it struggle to reach $M_\mathrm{dust}/M_\mathrm{stars}\sim10^{-2}$ even when assuming high supernova dust yields and a top-heavy IMF \citep[e. g.,][]{Burgarella2020DustConstraintsLbgs, Witstok2023DustEarlyEpochs}.

\subsection{Comparison to simulations}\label{sec:discussion:sims}

We compare our results with theoretical expectations by showing the predictions from three $z\approx2$ snapshots from cosmological simulations coupled with dust evolution models, namely two semi-analytic models (SAM) by \citet{Popping2017DustEvolutionSam} \citepalias{Popping2017DustEvolutionSam}, and  \citet{Parente2023CosmicDustSemiAnalytic} \citepalias{Parente2023CosmicDustSemiAnalytic}, and one hydrodynamical simulation with the MUlti Phase Particle Integrator (\textsc{MUPPI}) model \citep[][hereafter \citetalias{RagoneFigueroa2024MuppiSim}]{RagoneFigueroa2024MuppiSim}. Besides structural differences in the galaxy formation backbone of the simulations, all three models implement the main channels of dust creation, growth, and destruction, albeit with different recipes and approaches used for each process. 
For example, while \citetalias{Popping2017DustEvolutionSam} assumes a single dust grain size, \citetalias{Parente2023CosmicDustSemiAnalytic} and \citetalias{RagoneFigueroa2024MuppiSim} treat large and small grains separately. Unlike \citetalias{Popping2017DustEvolutionSam} and \citetalias{Parente2023CosmicDustSemiAnalytic}, \citetalias{RagoneFigueroa2024MuppiSim} also model H$_2$ formation in the surface of dust grains.  
Whereas for ISM dust growth, the three models compute accretion timescales in a similar way based on the \citet{Zhukovska2008IsmEvolutionStardust} approach, with an explicit dependence on metallicity, gas density, and the type of grain. 

The simulation snapshots provide catalogs with thousands of mock galaxies with a range of stellar mass, SFR, gas mass, and dust mass. For a consistent comparison, we restricted the stellar masses to $9<\log(M_\mathrm{stars}/M_\odot)<10.6$ matching the range spanned by our data. Ideally, we would also require $\mathrm{SFR} > \SI{20}{\msun\per\year}$ but imposing both selection cuts together leaves only 3\%, 0.7\%, and 0.8\% of the \citetalias{Parente2023CosmicDustSemiAnalytic}, \citetalias{RagoneFigueroa2024MuppiSim}, and \citetalias{Popping2017DustEvolutionSam} catalogs respectively. Matching $M_\mathrm{stars}$ and SFR would therefore make the catalogs too small to draw useful trends.

In the bottom left panel of \autoref{fig:dts_vs_oh} we see that the \citetalias{Parente2023CosmicDustSemiAnalytic} and \citetalias{RagoneFigueroa2024MuppiSim} models predict a flat or slightly decreasing trend for DtS against metallicity within the range of the ACE data, with both showing agreement on a  typical ratio $M_\mathrm{dust}/M_\mathrm{stars}\approx 10^{-2.7}$, which is 0.7 dex lower than the median observed value in the ACE sample.
On the other hand, the \citetalias{Popping2017DustEvolutionSam} simulation shows a DtS ratio that steadily grows from $10^{-2.6}$ to $10^{-2.3}$, between $\logoh=8.1$ and $\logoh=8.6$.
Again, our data can qualitatively reproduce this rising trend if we assume a metallicity-dependent $T_\mathrm{dust}$.


In the bottom right panel of \autoref{fig:dts_vs_oh}, the three models predict $M_\mathrm{dust}/\mathrm{SFR}$ values that are in excellent agreement with the ACE stacked observations at the observed metallicities, but only \citetalias{Parente2023CosmicDustSemiAnalytic} and \citetalias{Popping2017DustEvolutionSam} models reproduce a positive trend.
The \citetalias{RagoneFigueroa2024MuppiSim} predicts a flat trend. Here, assuming a metallicity-dependent $T_\mathrm{dust}$ steepens the relation, but also brings it further away from the model predictions.

The flat $M_\mathrm{dust}/\mathrm{SFR}$ behavior of the \citetalias{RagoneFigueroa2024MuppiSim} model can be attributed to a secondary dependence on stellar mass:  \citetalias{RagoneFigueroa2024MuppiSim} find a steep decline in $t_\mathrm{dep}$ as a function of stellar mass (see their Figure 12), and consequently as a function of metallicity via the MZR relation. This effectively counteracts the increase of \mdust and leads to a flat \mdust/SFR ratio in their model.

Of the three models considered here, \citetalias{Popping2017DustEvolutionSam} provides the best match overall to the ACE data in \autoref{fig:dts_vs_oh}.
However, since the three models have a similar treatment of dust physics, the differences likely arise from the galaxy formation scheme of each simulation.



\begin{figure*}[ht!]
	\resizebox{\hsize}{!}
	{\includegraphics {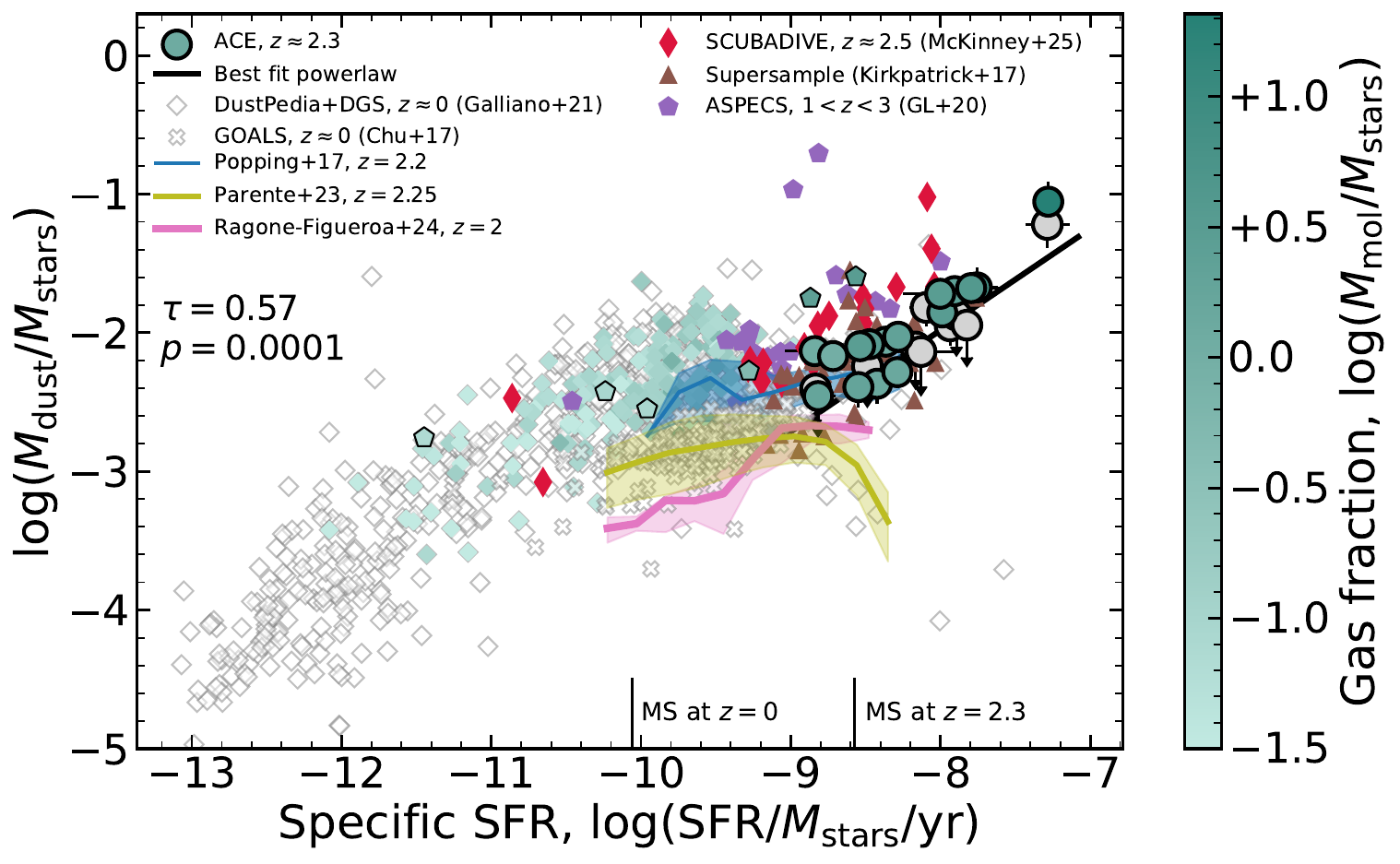}}
	\caption{
		Dust formation rate diagram (DtS versus specific SFR). 
		Galaxies from ACE are color coded by molecular gas fraction $M_\mathrm{mol}/M_\mathrm{stars}$ whenever they have a CO(3-2) detection \citep{Langan2026AceGfmr}, otherwise the marker is filled in gray.
		The same color code is applied to galaxies in the \citet{Galliano2021NearbyGalaxyDustEvol} sample that have CO-based $M_\mathrm{mol}$ measurements \citep{Casasola2020DustPediaMolGas}, and also to a handful of CO-detected ASPECS galaxies \citep{Aravena2020AspecsNatureOfFaintDusty}.
		The other markers are taken from \autoref{fig:mdustmulti} and \autoref{fig:dts_vs_oh}.
		The vertical lines at the bottom indicate the expected sSFR for an MS galaxy with $M_\mathrm{stars}=\SI{1e10}{\msun}$ at $z=0$ and $z=2.3$ according to \citet{Popesso2023MainSequenceAcrossCosmicTime}.
	}
	\label{fig:dts_vs_ssfr}
\end{figure*}

\subsection{Elevated dust-to-stellar mass ratio}

In Sections \ref{sec:analysis} and \ref{sec:discussion:nometaltrend} we showed that ACE galaxies have an increased DtS ratio with respect to local galaxies at the same stellar mass and metallicity.
However, the SFRs of ACE galaxies are also higher than their local counterparts.
In the literature, the excess of SFR at cosmic noon relative to $z=0$ is often attributed to the larger reservoirs of molecular gas available to form stars \citep[see][for a review]{Tacconi2020EvolutionISMReview}.
To illustrate the potential effect of SFR and gas fraction, we plot in \autoref{fig:dts_vs_ssfr} the DtS ratio against sSFR in the so-called ``dust formation rate'' diagram, with ACE galaxies color-coded by molecular gas fraction ($\mu_\mathrm{mol}=M_\mathrm{mol}/M_\mathrm{stars}$). 
The molecular gas masses were estimated from the CO(3-2) line (see \autoref{sec:obs:almaco}), assuming an excitation factor of $r_{31}=0.77$ \citep{Boogaard2020ASPECS-COExcitation} and the \citet{Accurso2017AlphaCO} metallicity-dependent $\alpha_\mathrm{CO}$, as described in detail in \citet{Langan2026AceGfmr}.
The resulting gas fractions go from $\log(\mu_\mathrm{mol})=-0.05$ to $\log(\mu_\mathrm{mol})=1.1$.
We also add $\mu_\mathrm{mol}$ information when available for the DustPedia sample \citep{Casasola2020DustPediaMolGas} and ASPECS \citep{GonzalezLopez2019Aspecs3mmCo} samples, showing that low sSFR galaxies have systematically lower $\mu_\mathrm{mol}$ than high sSFR galaxies at both epochs. 

We readily see that local galaxies follow a tight relation in the sSFR-DtS plane, as reported by several authors \citep[e. g.,][]{DaCunha2010PropertiesUlirgs, DeVis2017HerschelAtlasScalingRelations}.
In the ACE sample we recover a similar trend, characterized by a significant $\tau=0.57$ ($p=\num{7e-5}$), a powerlaw index of \num{0.73+-0.15}, and an intrinsic scatter of just $0.1$ dex (see \autoref{tab:regression}). We note, however, that DtS and sSFR are intrinsically correlated via \mstar by definition.

Regardless of their lower mass, lower metallicity, and rest-frame optical selection, ACE galaxies behave in a similar way to the IR-selected samples at cosmic noon. From massive ALMA-selected galaxies (ASPECS) to classical SMGs (SCUBADIVE, Supersample), the correlation between sSFR and DtS is present, as expected from numerous reports in the literature \citep[e. g.,][]{Martis2019DustMassiveGalaxiesUltraVista, Donevski2020DustToStellarEvolutionDsfg, Kokorev2021EvolvingIsmStardust}.

On the numerical side, the models introduced in \autoref{sec:discussion:sims} have very different predictions in this parameter space. The \citetalias{RagoneFigueroa2024MuppiSim} model exhibits an s-shaped increase with a transition at $\log(\mathrm{sSFR})\approx-9.2$, while \citetalias{Popping2017DustEvolutionSam} has high scatter at low sSFR but a clear DtS increase toward high sSFR. 
The \citetalias{Parente2023CosmicDustSemiAnalytic} model also shows  a rising trend but only at  $\log(\mathrm{sSFR})\lesssim-8.8$, after that it turns over to lower DtS ratios.
The \citetalias{Parente2023CosmicDustSemiAnalytic} curve resembles the behavior of analytic models with idealized star formation histories \citep[e. g.,][]{Asano2013DustFormationHistory, Nanni2020DustEvolLowMetallicity}.
In these models, the turnover in DtS can be interpreted as a manifestation of the critical metallicity: metal poor galaxies start a burst of star formation (high sSFR) at the bottom right of the diagram, where ISM dust growth is still inefficient. 
Right after they reach the critical metallicity, dust mass increases more rapidly than the rate of star formation, leading to a steep rise in DtS. Eventually they attain a maximum DtS while the starburst fades off, lowering the sSFR. Then, dust destruction overcomes production and the DtS decreases along with sSFR.

In this picture, the continued increase of DtS at high sSFR implied by the  $z\approx 2$ data plus the high $\mu_\mathrm{mol}$ values are consistent with the hypothesis that the ALMA-detected galaxies have a significant dust contribution from ISM growth. The data also suggest that the maximum DtS can be as high as $\log(M_\mathrm{dust}/M_\mathrm{stars})\approx -1$ provided there is a large enough reservoir of metal-enriched gas to fuel both star formation and ISM dust growth.

The general lack of agreement in this diagram between dust observations and simulations points to challenges on both sides. On one hand, populating the bottom right corner of the diagram requires deeper IR observations than currently possible at $z\approx 2$.
This region is likely occupied by starbursts with $\log(M_\mathrm{stars}/\mathrm{M}_\odot)\lesssim8$. Then, getting at  $\log(M_\mathrm{dust}/M_\mathrm{stars})<-3$ would mean detecting a dust mass of less than \SI{1e5}{\msun}, which is more than two orders of magnitude below our detection limit.
On the other hand, simulations generally struggle to produce galaxies with high sSFR that are as massive as those of the ACE sample.

\subsection{Caveats}\label{sec:discussion:caveats}
The main uncertainty of this study comes from the lack of robust galaxy-to-galaxy constraints on $T_\mathrm{dust}$, $\beta$, and $\kappa_0$. As described in \autoref{sec:analysis:dustmass}, we based our assumptions on data and previous literature on the topic, however, as the dust masses are derived from a single FIR band, the assumptions can introduce uncertainties on a galaxy-to-galaxy basis.
Additionally, we have assumed no evolution of these quantities across the redshift range considered here.

In particular, we assumed MW-like dust properties with a fixed cross section normalization $\kappa_0$.
Theoretically $\kappa_0$ depends on the grain composition and size distribution, therefore we could expect a metallicity or redshift dependence.
But empirical constraints on $\kappa_0$ are difficult to obtain. For example, \citet{Bianchi2019Dustpedia} used the gas mass, metallicity, and FIR observations of DustPedia galaxies to indirectly infer $\kappa_0$ maps.
The study finds a wide range of $\kappa_0$ values across the sample, and a mild trend with metallicity.
However, the results are still based on assumed prescriptions of DtM and DtG fractions as function of metallicity.

The second limitation of this study is the sample size. While ACE uniquely expands our constraints on dust mass to lower metallicities and lower stellar masses than previously explored at this redshift, larger samples at these metallicities are needed to place more robust constraints on scaling relations.  With the growing number of spectroscopic samples in JWST deep fields, it will soon be possible to use ALMA to target a much larger number of galaxies with metallicities even lower than those of ACE, albeit with a significant time investment. More efficient approaches would involve wide  field surveys with future facilities such as the proposed Atacama Large Aperture Submillimeter Telescope \citep[AtLAST; e. g.,][]{vanKampen2024AtlastHighzSurvey} or more sensitive interferometers \citep[e. g., as envisioned by the ALMA2040 concept,][]{Algera2026Alma2040DustCosmicDawn}.

\section{Summary and conclusions}\label{sec:conclusions}

In this work, we have analyzed the dust mass content of a sample of 25 star-forming galaxies of subsolar metallicity at $z\approx2.3$ using ALMA data from the ACE Large Program. The sample is a subset of the MOSDEF survey, enabling robust metallicities and SFR diagnostics. The main results from our work can be summarized as follows:
\begin{itemize}
	\item Under the assumption of a fixed dust temperature $T_\mathrm{dust}=\SI{25}{\kelvin}$,  we find positive trends for dust mass as a function of stellar mass, metallicity, and SFR. At face value, \mdust shows the strongest correlation with SFR, and is consistent with following a powerlaw with a linear slope (=\num{1.0+-0.3}). We discuss this result in the light of the evolutionary framework proposed by \citet{Hjorth2014DustSfrRelation}.
	\item We find that the \mdust/\mstar (DtS) ratio scatters around $10^{-2}$ ($4-9\times$ higher than local samples), and shows no significant correlation with metallicity down to $0.3Z_\odot$, while the \mdust/SFR ratio increases with metallicity characterized by a powerlaw slope of \num{1.6+-0.5}. The latter mimics the behavior of the $M_\mathrm{dust}/M_\mathrm{mol}$ ratio reported in the companion paper of \citet{Geesink2026AceGasToDustRatio} since the the gas depletion timescale is found to be constant with metallicity in the ACE sample \citep{Popping2026AceGasScalingRelations}. The DtS and \mdust/SFR  trends become positively steeper if we assume an inverse relationship between $T_\mathrm{dust}$ and metallicity, instead of a fixed temperature. In any case, the results are in broad qualitative agreement with cosmological simulations, especially with the \citetalias{Popping2017DustEvolutionSam} model,  and are consistent with a significant contribution of ISM growth to the dust buildup.
	\item Lastly, we find a correlation between specific dust mass $M_\mathrm{dust}/M_\mathrm{stars}$ and specific SFR (sSFR).
		This correlation has been reported extensively for regular galaxies in the local Universe, and for massive dusty galaxies at $z>1$. Here, we find the correlation still holds in subsolar-metallicity, yet highly star-forming ($\mathrm{sSFR}>\SI{1e-8}{\per\year}$) galaxies at cosmic noon.
	This finding also supports the hypothesis of a strong contribution of ISM growth, since we do not observe an anticorrelation or turnover at high sSFR as predicted by some models. Finally, we look at the molecular gas masses inferred from ACE's CO(3-2) observations and argue that the large molecular gas fractions ($M_\mathrm{mol}/M_\mathrm{stars}\gtrsim 1$) observed in our sample drive both the high sSFR and the high DtS ratio.

\end{itemize}

While these results are dependent on a series of assumptions on the intrinsic properties of dust, they provide new constraints on the dust mass scaling relations at $z\approx 2$, and therefore shed light on the lifecycle of dust and its interplay with star formation.

\begin{acknowledgements}
The authors wish to thank Hiddo Algera and Tom Bakx for helpful discussions in relation to this work.
This work has been funded by the European Research Council (ERC) under the European Union's Horizon 2020 research and innovation program (DistantDust, Grant agreement No. 101117541). IS and IL also acknowledge funding from the Atracci\'on de Talento Grant No. 2022-T1/TIC-20472 of the Comunidad de Madrid, Spain. LA acknowledges support by the CSIC Program 'Programa JAE' (JAE-Pre 2023), by the grant PID2024-158856NA-I00 funded by Spanish Ministerio de Ciencia e Innovación MCIN/AEI/10.13039/501100011033 and by “ERDF A way of making Europe” in addition to the aforementioned ERC grant.
L.A.B. acknowledges support from the Dutch Research Council (NWO) under grant VI.Veni.242.055 (\url{https://doi.org/10.61686/LAJVP77714}).DN is grateful for support from NASA via grants ATP-21-0013 and ATP-23-0002. MP is funded by NASA grant ATP-23-0002.
This paper makes use of the following ALMA data: ADS/JAO.ALMA\#2018.1.01128.S, 2019.1.01142.S, 2024.1.00534.L. ALMA is a partnership of ESO (representing its member states), NSF (USA) and NINS (Japan), together with NRC (Canada), MOST and ASIAA (Taiwan), and KASI (Republic of Korea), in cooperation with the Republic of Chile.
We acknowledge assistance and computational support provided by Allegro, the European ALMA Regional Center node in the Netherlands.
\end{acknowledgements}

%
\bibliographystyle{aa}
\bibliography{refs.bib}

\begin{appendix}
\onecolumn
\section{Dust masses and other properties of the ACE sample}\label{sec:ap:dustmass}
In this section we include the main table of this paper, listing the dust mass of every ACE galaxy along with other properties and identifiers.

\begingroup
\renewcommand*{\arraystretch}{1.5}
\begin{table}[ht!]
	\caption{Dust masses and other properties of the ACE sample.}	
	\label{tab:mdust}
	\centering
	\begin{tabular}{cccccccccc}
\hline\hline
 (1) & (2) & (3) & (4) & (5) & (6) & (7) & (8) & (9) & (10) \\
ID & R. A. & Dec. & $z$ & $M_\mathrm{stars}$ & SFR & $12+\log(\mathrm{O}/\mathrm{H})$ & $\nu_\mathrm{obs}$ & $S_\nu$ & $M_\mathrm{dust}$ \\
 & deg & deg & & \SI{1e9}{\msun} & \si{\msun\per\year} & & \si{\giga\hertz} & \si{\micro\jansky} & \SI{1e7}{\msun}   \\
\hline
3324 & 150.14841 & 2.21313 & 2.307 & $19.9_{-3.2}^{+4.1}$ & $86 \pm 37$ & $8.57 \pm 0.05$ & 343.5 & $319\pm44$ & $17.1_{-6.2}^{+14.6}$\\
3626 & 150.10474 & 2.21573 & 2.325 & $14.7_{-1.6}^{+1.5}$ & $22 \pm 2$ & $8.19 \pm 0.03$ & 343.5 & $<108$ & $<5.8$\\
3666 & 150.07753 & 2.21603 & 2.086 & $1.6_{-0.2}^{+0.4}$ & $84 \pm 10$ & $8.29 \pm 0.02$ & 343.5 & $259\pm71$ & $14.1_{-5.5}^{+12.0}$\\
3773 & 150.19815 & 2.21659 & 2.425 & $4.4_{-0.7}^{+0.8}$ & $56 \pm 15$ & $8.27 \pm 0.03$ & 233 & $<34$ & $<5.8$\\
4497 & 150.07147 & 2.22387 & 2.441 & $17.9_{-4.1}^{+4.4}$ & $68 \pm 20$ & $8.51 \pm 0.04$ & 343.5 & $143\pm35$ & $7.7_{-3.1}^{+7.1}$\\
5094 & 150.14037 & 2.23019 & 2.171 & $21.3_{-9.9}^{+7.7}$ & $71 \pm 24$ & $8.47 \pm 0.05$ & 343.5 & $326\pm49$ & $17.6_{-6.3}^{+14.5}$\\
5814 & 150.16913 & 2.23840 & 2.127 & $21.9_{-2.3}^{+2.7}$ & $114 \pm 17$ & $8.45 \pm 0.02$ & 343.5 & $378\pm61$ & $20.5_{-7.4}^{+16.9}$\\
5901 & 150.18941 & 2.23814 & 2.396 & $5.3_{-0.7}^{+1.6}$ & $42 \pm 6$ & $8.45 \pm 0.03$ & 343.5 & $150\pm36$ & $8.0_{-3.2}^{+7.3}$\\
6283 & 150.11528 & 2.24191 & 2.224 & $7.1_{-1.6}^{+1.6}$ & $49 \pm 5$ & $8.22 \pm 0.03$ & 343.5 & $<104$ & $<5.6$\\
6750 & 150.16302 & 2.24749 & 2.127 & $4.7_{-0.6}^{+0.6}$ & $55 \pm 6$ & $8.35 \pm 0.02$ & 343.5 & $98\pm29$ & $5.3_{-2.2}^{+4.6}$\\
8280 & 150.08650 & 2.26432 & 2.494 & $39.0_{-2.0}^{+2.0}$ & $56 \pm 26$ & $8.51 \pm 0.05$ & 343.5 & $536\pm49$ & $28.7_{-10.3}^{+25.0}$\\
8515 & 150.18448 & 2.26626 & 2.454 & $1.3_{-0.2}^{+0.6}$ & $67 \pm 11$ & $8.27 \pm 0.03$ & 233 & $46\pm12$ & $7.8_{-2.8}^{+5.7}$\\
9393 & 150.16354 & 2.27509 & 2.413 & $7.8_{-1.4}^{+2.0}$ & $80 \pm 11$ & $8.37 \pm 0.02$ & 343.5 & $203\pm51$ & $10.9_{-4.5}^{+10.3}$\\
9971 & 150.14354 & 2.28180 & 2.411 & $25.6_{-4.5}^{+5.3}$ & $72 \pm 7$ & $8.36 \pm 0.02$ & 343.5 & $193\pm53$ & $10.4_{-4.3}^{+9.3}$\\
13296 & 150.11510 & 2.31529 & 2.167 & $18.1_{-2.2}^{+2.2}$ & $35 \pm 13$ & $8.61 \pm 0.05$ & 343.5 & $227\pm51$ & $12.3_{-4.7}^{+10.4}$\\
13701 & 150.11272 & 2.31944 & 2.166 & $13.2_{-0.7}^{+1.8}$ & $165 \pm 34$ & $8.55 \pm 0.03$ & 343.5 & $489\pm57$ & $26.4_{-9.3}^{+22.4}$\\
16594 & 150.12495 & 2.35022 & 2.286 & $9.2_{-1.9}^{+1.9}$ & $27 \pm 6$ & $8.45 \pm 0.03$ & 343.5 & $137\pm35$ & $7.4_{-3.0}^{+6.5}$\\
19013 & 150.11200 & 2.37263 & 2.457 & $7.5_{-1.2}^{+1.2}$ & $74 \pm 41$ & $8.52 \pm 0.03$ & 343.5 & $265\pm43$ & $14.2_{-5.5}^{+12.7}$\\
19439 & 150.10150 & 2.37672 & 2.466 & $5.6_{-1.0}^{+1.1}$ & $84 \pm 13$ & $8.22 \pm 0.02$ & 343.5 & $<118$ & $<6.3$\\
19985 & 150.06035 & 2.38277 & 2.188 & $28.3_{-3.5}^{+3.5}$ & $146 \pm 9$ & $8.31 \pm 0.01$ & 343.5 & $272\pm35$ & $14.7_{-5.0}^{+12.3}$\\
21955 & 150.09535 & 2.40281 & 2.468 & $4.2_{-0.9}^{+1.0}$ & $73 \pm 17$ & $8.50 \pm 0.03$ & 343.5 & $165\pm39$ & $8.8_{-3.6}^{+8.5}$\\
22193 & 150.08542 & 2.40596 & 2.465 & $7.8_{-1.3}^{+3.3}$ & $58 \pm 26$ & $8.44 \pm 0.04$ & 343.5 & $<106$ & $<5.7$\\
24020 & 150.11517 & 2.42553 & 2.092 & $8.9_{-1.7}^{+1.5}$ & $29 \pm 9$ & $8.32 \pm 0.03$ & 343.5 & $<94$ & $<5.1$\\
24763 & 150.05670 & 2.43466 & 2.464 & $9.0_{-1.3}^{+1.4}$ & $145 \pm 75$ & $8.45 \pm 0.05$ & 343.5 & $354\pm50$ & $19.0_{-7.2}^{+17.3}$\\
25229 & 150.10841 & 2.43971 & 2.181 & $15.9_{-2.7}^{+2.8}$ & $24 \pm 5$ & $8.29 \pm 0.03$ & 343.5 & $<102$ & $<5.5$\\
\hline
\end{tabular}

	\tablefoot{
		(1) Source number from the 3D-HST catalog version 4 \citep{Skelton2014Hst3dCatalog}. (2) Right Ascension (J2000). (3) Declination (J2000).
		(4). Optical redshift from MOSFIRE.
		(5). Stellar mass estimated with \textsc{prospector}.
		(6). \halpha-based SFR with reddening correction from \halpha/\hbeta ratio \citep{Shivaei2022InfraredSedSubsolar}.
		(7). Gas-phase nebular oxygen abundance from strong-line ratios with the \citet{Sanders2026AuroraMetallicity} calibration.
		(8). Observed-frame frequency $\nu_\mathrm{obs}$ for dust mass estimation.
		Possible values are \SI{343.5}{\giga\hertz} (Band 7) and \SI{233}{\giga\hertz} (Band 6).
		(9). Continuum flux density at $\nu_\mathrm{obs}$. Values prefixed with $<$ represent $3\sigma$ upper limits.
		(10). Single-band, optically thin MBB dust mass (\autoref{eq:dmass}) derived from either ALMA Band 7 or 6 data.
		The calculation assumes $T_\mathrm{dust}=\SI{25}{\kelvin}$, $\beta=2.08$, and $\kappa_0 = \SI{0.4}{\meter\squared\per\kilogram}$ at $\lambda_0=\SI{250}{\micro\meter}$ \citep{Draine2003InterstellarDustGrainsReview}.
		The uncertainties include the flux measurement error and a systematic $\sigma=\SI{5}{\kelvin}$ uncertainty in the dust temperature.
	}
\end{table}
\endgroup

\section{Stacks}\label{sec:ap:stacks}

In addition to the individual dust masses, we performed measurements on stacked data of subsamples within ACE.
We split the sample into pre-defined metallicity and stellar mass bins, as specified in \autoref{tab:stacks}.
For each bin, we stacked the ALMA \ang{;;2}-tapered images with a common restoring beam and combined them through an inverse-rms weighted mean pixel by pixel.
The flux densities and uncertainties were computed through a combined Monte Carlo and $N-1$ bootstrap scheme that accounts for intrinsic measurement uncertainty, bin assignment uncertainty, and the sensitivity to which set of galaxies  are included in each stack. Briefly, each bin is populated by randomly sampling each galaxy's \logoh~or $M_\mathrm{stars}$ from its associated uncertainty. At the same time, we randomly remove one galaxy per bin in each realization and stack the remaining sources. This process is repeated 500 times, resulting in a distribution of flux densities, from which we report its median and 16th and 84th percentiles. 
Further details on the binning method are provided in \citet{Geesink2026AceGasToDustRatio}.

For bins with detections in both Band 6 and Band 7, we computed dust masses by ``fitting'' MBBs with all parameters fixed except $M_\mathrm{dust}$.
The fixed parameters are the same specified in \autoref{sec:analysis:dustmass}.
The maximum likelihood $M_\mathrm{dust}$ was then obtained with SciPy's \verb|minimize| routine. The fit was repeated 1000 times with different fixed temperatures drawn randomly from a normal distribution centered at \SI{25}{\kelvin} and standard deviation of \SI{5}{\kelvin}, so that the 16\textsuperscript{th}-84\textsuperscript{th} percentiles of the resulting $M_\mathrm{dust}$ distribution capture the uncertainty on the assumed dust temperature.
For the bins with only a detection in Band 7, we computed a single-band $M_\mathrm{dust}$ following the procedure outlined in \autoref{sec:analysis:dustmass}.

\begingroup
\renewcommand*{\arraystretch}{1.5}
\begin{table}[!htb]
  \centering
  \caption{Stacked ACE data in bins of metallicity and stellar mass.}
  \label{tab:stacks}
  \begin{tabular}{cccc}
    \hline\hline
   $\logoh$ & $<8.38$ & $8.38-8.48$ & $>8.48$ \\
   \hline
   	\# of sources & 12  & 6 & 7 \\
	$z_\mathrm{spec}$ & $2.27_{-0.05}^{+0.05}$ & $2.40_{-0.08}^{+0.03}$ & $2.37_{-0.07}^{+0.08}$ \\
	$\log(M_\mathrm{stars}/\mathrm{M}_\odot)$ & $9.86_{-0.09}^{+0.07}$ & $9.99_{-0.10}^{+0.15}$ & $10.18_{-0.08}^{+0.07}$ \\
	$\log(\mathrm{SFR}/\mathrm{M}_\odot/\mathrm{yr})$ & $1.78_{-0.06}^{+0.05}$ & $1.80_{-0.15}^{+0.11}$ & $1.85_{-0.10}^{+0.11}$ \\
	$S_{873\mu\mathrm{m}}/\mu\mathrm{Jy}$ & $91_{-15}^{14}$ & $198_{-42}^{+47}$ & $271_{-33}^{+34}$ \\
	$S_{1287\mu\mathrm{m}}/\mu\mathrm{Jy}$ & $31_{-5}^{+5}$ & $<60$  & $76_{-18}^{+17}$\\
	$\log(M_\mathrm{dust}/\mathrm{M}_\odot)$ & $7.70_{-0.17}^{+0.26}$ & $8.02_{-0.18}^{+0.27}$ & $8.14_{-0.16}^{+0.24}$\\
    \hline\hline
	$\log(M_\mathrm{stars}/\mathrm{M}_\odot)$ & $<9.80$ & $9.8-10.2$ & $>10.2$ \\
	\hline
	\# of sources & 7 & 9 & 9 \\
	$z_\mathrm{spec}$ & $2.43_{-0.03}^{+0.01}$ & $2.32_{-0.04}^{+0.08}$ & $2.25_{-0.06}^{+0.06}$ \\
	$12+\log(\mathrm{O}/\mathrm{H})$ & $8.31_{-0.03}^{+0.04}$ & $8.42_{-0.04}^{+0.03}$ & $8.46_{-0.03}^{+0.03}$ \\
	$\log(\mathrm{SFR}/\mathrm{M}_\odot/\mathrm{yr})$ & $1.81_{-0.06}^{+0.06}$ & $1.71_{-0.11}^{+0.13}$ & $1.87_{-0.10}^{+0.07}$\\
	$S_{873\mu\mathrm{m}}/\mu\mathrm{Jy}$ & $96_{-20}^{+23}$ & $143_{-24}^{+28}$ & $257_{-27}^{+35}$ \\
	$S_{1287\mu\mathrm{m}}/\mu\mathrm{Jy}$ & $36_{-10}^{+7}$ & <22 & $70_{-13}^{+29}$\\
	$\log(M_\mathrm{dust}/\mathrm{M}_\odot)$ & $7.75_{-0.19}^{+0.25}$ & $7.90_{-0.18}^{+0.27}$ & $8.12_{-0.17}^{+0.20}$\\
	\hline
  \end{tabular}	
  \tablefoot{The first half of the table contains the stacked results in three metallicity bins, while the second half is split in three stellar mass bins. Flux values prefixed with $<$ represent $3\sigma$ upper limits.}
\end{table}
\endgroup

\section{Metallicity-dependent dust temperature}\label{sec:ap:tdz}
In this section, we build on the results from MBB fitting to composite ACE SEDs presented in Popescu et al. in prep., and recompute dust masses with a metallicity-based prescription for the dust temperature. A linear fit to the three binned points resulted in a slope of $-66.67\,\mathrm{K}/\mathrm{dex}$. We normalized the relation so that the highest metallicity bin ($\logoh=8.55$) matches our fiducial $T_\mathrm{dust}=\SI{25}{\kelvin}$:

\begin{equation}\label{eq:tdz}
	T_\mathrm{dust}/\mathrm{K} = -66.67 \times (12 + \log(\mathrm{O}/\mathrm{H})) + 595.
\end{equation}

\begin{figure}[!hbt]
	\includegraphics[width=\linewidth]{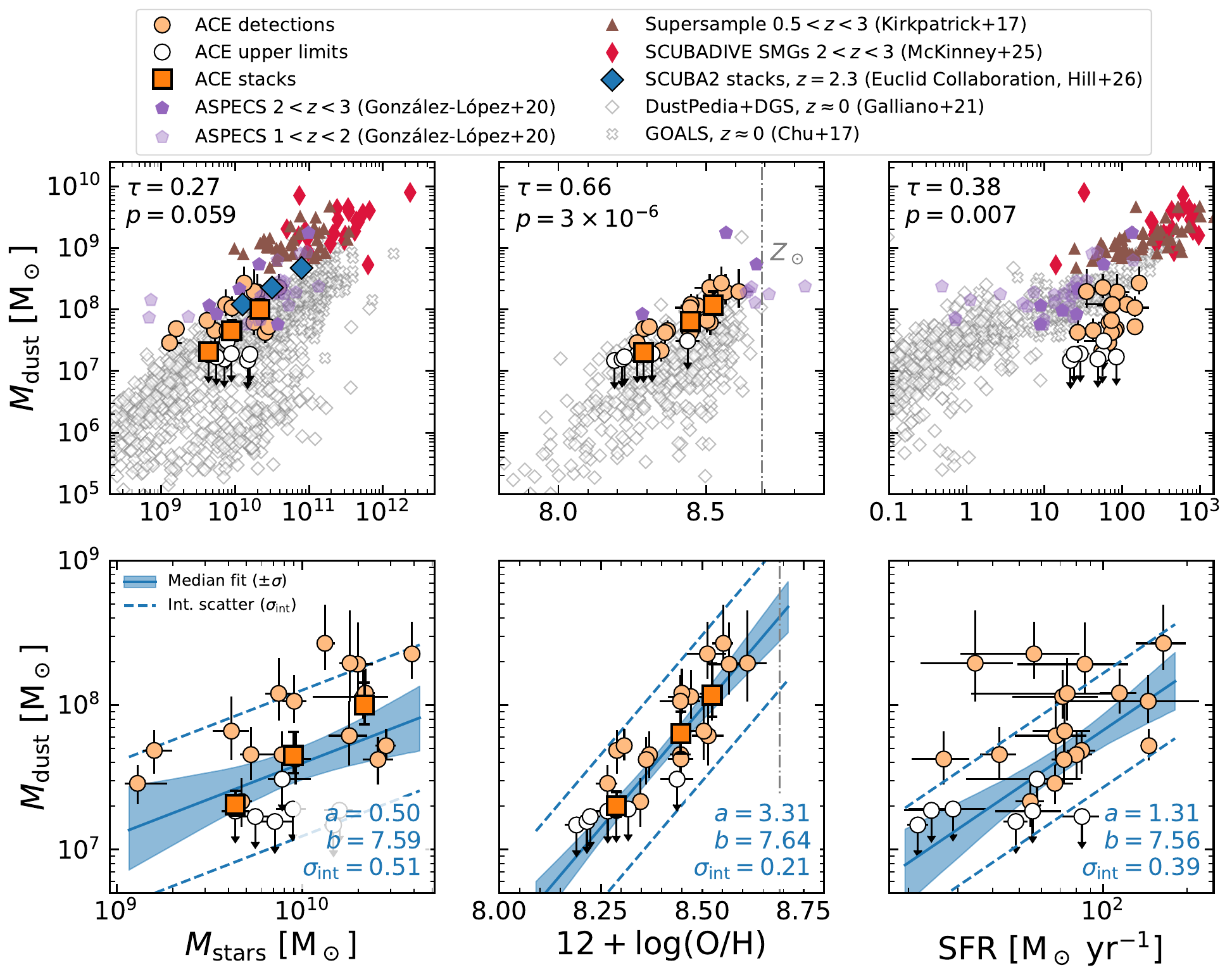}
	\caption{Same as \autoref{fig:mdustmulti} but with metallicity dependent $T_\mathrm{dust}$ according to \autoref{eq:tdz}.}\label{fig:mdustmultitdz}
\end{figure}

In \autoref{fig:mdustmultitdz} we remade \autoref{fig:mdustmulti} but with dust masses for ACE calculated with \autoref{eq:tdz}. The \mdust-metallicity relation (middle column)) steepens to a powerlaw index of $3.3$ while the other relations significantly increase in scatter, and the \mdust$-M_\mathrm{stars}$ relation fails to pass the correlation test with $p>0.05$. All the other correlation tests and linear regression results are shown in \autoref{tab:regressionTdZ}.

\begingroup
\renewcommand*{\arraystretch}{1.5}
\begin{table*}[!hbt]
  \centering
  \caption{Results of the correlation test and linear regression described in \autoref{sec:analysis:scaling} (same as \autoref{tab:regression}) but applied to the $T_\mathrm{dust}(Z)-$derived dust masses.}
  \label{tab:regressionTdZ}
  \begin{tabular}{cccccccc}
  \hline\hline
  $x$ & $y$ & $\tau$ & $p$ &$\log(x_0)$ & $a$ & $b$ & $\sigma_\mathrm{int}$ \\
  \hline
  $M_\mathrm{stars}$ & $M_\mathrm{dust}$ & 0.27 & 0.059 & 9.98 & $0.50_{-0.30}^{+0.28}$ & $7.59_{-0.11}^{+0.11}$ & $0.51_{-0.09}^{+0.13}$\\
$12+\log(\mathrm{O}/\mathrm{H})$ & $M_\mathrm{dust}$ & 0.66 & \num{3.4e-06} & 8.40 & $3.31_{-0.50}^{+0.51}$ & $7.64_{-0.06}^{+0.06}$ & $0.21_{-0.06}^{+0.08}$\\
$\mathrm{SFR}$ & $M_\mathrm{dust}$ & 0.38 & 0.007 & 1.79 & $1.31_{-0.42}^{+0.45}$ & $7.56_{-0.11}^{+0.09}$ & $0.39_{-0.09}^{+0.12}$\\
$12+\log(\mathrm{O}/\mathrm{H})$ & $M_\mathrm{dust}/M_\mathrm{stars}$ & 0.40 & 0.005 & 8.40 & $2.32_{-0.78}^{+0.84}$ & $-2.39_{-0.11}^{+0.10}$ & $0.45_{-0.08}^{+0.12}$\\
$12+\log(\mathrm{O}/\mathrm{H})$ & $M_\mathrm{dust}/\mathrm{SFR}$ & 0.68 & \num{1.7e-06} & 8.40 & $2.99_{-0.43}^{+0.47}$ & $5.80_{-0.05}^{+0.05}$ & $0.08_{-0.05}^{+0.07}$\\
$\mathrm{sSFR}$ & $M_\mathrm{dust}/M_\mathrm{stars}$ & 0.36 & 0.012 & -8.19 & $0.73_{-0.22}^{+0.24}$ & $-2.40_{-0.09}^{+0.09}$ & $0.39_{-0.09}^{+0.12}$

  \\ \hline
  \end{tabular}
\end{table*}
\endgroup

\end{appendix}

\end{document}